\documentclass[10pt,letterpaper]{article}

\DeclareMathAlphabet{\mathpzc}{OT1}{pzc}{m}{it}

\usepackage{mathrsfs}

\usepackage{graphicx}

\usepackage{amssymb}

\usepackage[figuresright]{rotating}

\usepackage{float}

\usepackage{amsmath}

\usepackage{amsfonts}

\usepackage{theorem}
\theorembodyfont{\upshape}

\newcommand{\tableline}{\hline }

\newcommand{\be}{\begin{equation}} 
\newcommand{\ee}{\end{equation}}
\newcommand{\bdm}{\begin{displaymath}} 
\newcommand{\edm}{\end{displaymath}}
\newcommand{\bea}{\begin{eqnarray}} 
\newcommand{\eea}{\end{eqnarray}}
\newcommand{\bse}{\small \begin{equation}} 
\newcommand{\ese}{\end{equation} \normalsize}
\newcommand{\bsea}{ \small \begin{eqnarray}} 
\newcommand{\esea}{\end{eqnarray} \normalsize}
\newcommand{\nn}{\nonumber \\} 

\newcommand{\fig}{Fig.~\ref} 
\newcommand{\tab}{Table~\ref}
\newcommand{\sect}{Section~\ref}
\newcommand{\eqn}{Eq.~\ref}

\newcommand{\bi}{{\mathscr{B}}}
\newcommand{\con}{{\mathscr{C}}}

\begin{document}
\title{{Ninth and Tenth Order Virial Coefficients for Hard Spheres in $D$ Dimensions}}

\author{Nathan~Clisby\footnote{e-mail:\nobreak{N.Clisby@ms.unimelb.edu.au}}
  \\ARC Centre of Excellence for Mathematics and Statistics of Complex
  Systems\\ 139 Barry Street\\ The University of Melbourne, Parkville Victoria 3010\\ Australia \and Barry~M.~McCoy\footnote{e-mail:\nobreak{mccoy@insti.physics.sunysb.edu}}\\C.~N.~Yang Institute for Theoretical Physics\\ Stony Brook University\\ Stony Brook, NY 11794-3840}

\maketitle

\begin{abstract}
We evaluate the virial coefficients $B_k$ for $k\leq 10$ 
for hard spheres in dimensions $D=2,\cdots ,8.$
Virial coefficients with $k$ even are found to be negative when 
$D\geq 5. $  This provides strong evidence
that the leading singularity for the virial series lies away from the
positive real axis when $D\geq 5$. Further analysis provides evidence that 
negative 
virial coefficients will be seen for some $k>10$  for $D=4$, and
there is a distinct possibility that negative virial coefficients
will also eventually occur for $D=3.$ 
\end{abstract}


\section{Introduction}
\label{intro}

The hard sphere gas of particles of diameter $\sigma$
in $D$ dimensions defined
by the two body
potential
\be
U({\bf r}) = \left\{ \begin{array}{c} +\infty \\ 0 \end{array}
\right. \begin{array}{c} |{\bf r}|<\sigma \\ |{\bf r}|> \sigma \\
\end{array}
\label{eq:hardspherepotential}
\ee
is one of the oldest and most studied systems in statistical mechanics.
Nevertheless after more than a century after the initial work of
van~der~Waals~\cite{vanderwaals1899a},
Boltzmann~\cite{boltzmann1899b}, and van~Laar~\cite{vanlaar1899b}
who computed the coefficients
up through $B_4$ in three dimensions for the low density virial
expansion of the pressure
\be \frac{P}{k_BT}=
\sum_{k=1}^{\infty}B_k \rho^{k} \hspace{0.5cm} \mathrm{with} \; B_1\equiv 1
\label{eq:virial}
\ee
there are still basic features of this system which are unresolved
and controversial. Chief among these properties are 1)
the radius of convergence of the virial series, 2) the question of
the occurrence of negative virial coefficients in 2 and 3 dimensions, and 3)
the analytic demonstration of the freezing phase transition which was
first seen in computer experiments~\cite{alder1957a,wood1957a} in the late 1950s.

The analytic information available for the hard sphere gas is extremely
sparse. Indeed, outside of $B_2$ and $B_3$ which are elementary to compute
and the original computation of $B_4$ in $D=3$ by Boltzmann and van~Laar, the
only other analytic computations of virial coefficients are of $B_4$
in $D=2$ done simultaneously by Rowlinson~\cite{rowlinson1964a} and
Hemmer~\cite{hemmer1964a} in 1964, and the recent computation
by the present authors~\cite{clisby2004a} of $B_4$ for
$D=4,6,8,10,12$, and by Lyberg~\cite{lyberg2004a} for odd dimensions
$D=5,7,9,11$.

All other computations for the hard sphere gas are by means of computer.
The fifth virial coefficient was first calculated numerically in the
1950s for hard discs by Metropolis et al.~\cite{metropolis1953a} and
for hard spheres by Rosenbluth and
Rosenbluth~\cite{rosenbluth1954a}. An extensive and systematic effort
to numerically calculate 
the virial coefficients $B_5$, $B_6$, and
$B_7$ for two and three dimensions was carried out by Ree and
Hoover\cite{ree1964a,ree1964c,ree1967a} during the 1960s.
The coefficient $B_8$ in
$D=2$ and $3$ was computed by Janse van
Rensburg~\cite{jansevanrensburg1993a} in 1993. Vlasov, You, and
Masters~\cite{vlasov2002a} recently 
recalculated $B_7$ and $B_8$, and Lab\'ik, Kolafa, and
Malijevsk\'y~\cite{kolafa2004a,labik2004a} have recalculated the
virial coefficients to $B_8$ and also calculated $B_9$.
All these virial coefficients are positive.

The study of virial coefficients for dimensions greater than 3 was
initiated in 1964 by Ree and Hoover~\cite{ree1964b} who computed $B_4$
for $D=4,\cdots,11$ and found for $D \geq 8$ that $B_4$ is negative.
This was the first time that any negative virial coefficient had been
seen for the hard sphere system. The coefficients $B_5$ and
$B_6$ for $D=4$ and $5$ were computed by Bishop, Masters, and
Clarke~\cite{bishop1999a} in 1999, and Bishop, Masters, and
Vlasov~\cite{bishop2004a} have recently calculated $B_7$ in dimensions
four and five, and $B_8$ in four dimensions. The present 
authors recently extended~\cite{clisby2004b} the study of $B_4$, $B_5$, and $B_6$
up to dimensions $D=50$.
In this study it was found that $B_5$ is always positive but the $B_6$
is negative for $D\geq 6$.

There are two very important observations to be made about these computations.

The first observation is that the virial coefficients
$B_2$ through $B_8$ in $D=2,3$ have been analyzed
by many authors in an attempt to find an approximate equation of
state. These approximations are reviewed in Subsection \ref{scenarios}.
Without exception all of these studies have the remarkable
feature that they show no singularity at the density at which the
computer studies indicates that the system freezes. Indeed some of the
approximate equations of state are analytic for real positive densities
greater than the density of closest packed discs and spheres.
One interpretation of this is that the first eight virial coefficients are not
sufficient to see the true asymptotic behavior of $B_k$ for large $k$.

The second observation is that because $B_4$ changes sign between $D=7$ and
$D=8$, and $B_6$ changes sign between $D=5$ and $6$ it may
be for large $k$ that $B_k$ can become negative for dimensions smaller than 5.
In particular if for $D=2$ or $D=3$ there were a value of $k$ such that
$B_k$ changed sign then the approximate
equations of state obtained from the first eight virial coefficients
would be wholly inadequate to obtain the radius of convergence 
of the virial series.

In this paper we address the questions of the radius of convergence and
and the negativity of the virial coefficients by numerically 
computing $B_7$ for $D=6,7,8$, $B_8$ for $D=5,\cdots,8$,
and $B_9$ and $B_{10}$ for $D=2,\cdots,8$. Our results are given
in \tab{tab:numericalvirial}.
For the Monte-Carlo integration we use the formulation of Ree and
Hoover~\cite{ree1964a,ree1964c}. For hard spheres it is well
known that in any dimension $D$ there may be some Ree-Hoover diagrams
which vanish identically for geometric reasons. The determination of the
number of such geometrically excluded diagrams is an interesting problem
in its own right and our results for 
this are given in \tab{tab:diagram_number}.

In \sect{virial} we discuss the virial expansion and the 
theoretical framework for the problem of calculating 
virial coefficients for hard spheres.
In \sect{method} we explain our method of computation with particular
emphasis on what needs to be done to handle the 4,980,756 Ree-Hoover
diagrams contributing to $B_{10}$ in an efficient fashion. In
\sect{background} we provide information about the hard sphere system
relevant to the understanding of the asymptotic behavior of the
virial series.
In \sect{discussion} we address the question of the large $k$ behavior
of $B_k.$  We propose in Subsection \ref{zero} two complementary
criteria which $B_k$ 
should satisfy in order to be considered asymptotic. We discuss the
asymptotic behavior of the virial series in \ref{ratio} from a ratio
analysis.
In Subsection \ref{diff} we apply the methods of Pad\'e approximants
and differential approximants to the ten known coefficients.
We conclude in \sect{conclusion} with a summary of key results.

The results of this work as well as the papers of 
Clisby and McCoy~\cite{clisby2004a,clisby2004b} are included 
in the dissertation of Clisby~\cite{clisby2004c}.

\begin{sidewaystable}
\centering
\scriptsize
\begin{tabular}{lllllllll}
\tableline
\tableline
\\[-1.5ex]
\multicolumn{1}{c}{$D$} &\multicolumn{1}{c}{$B_3/B_2^2$}
&\multicolumn{1}{c}{$B_4/B_2^3$} &\multicolumn{1}{c}{$B_5/B_2^4$}
&\multicolumn{1}{c}{$B_6/B_2^5$} &\multicolumn{1}{c}{$B_7/B_2^6$}
&\multicolumn{1}{c}{$B_8/B_2^7$} &\multicolumn{1}{c}{$B_9/B_2^8$}
&\multicolumn{1}{c}{$B_{10}/B_2^9$} \\[0.7ex]
\tableline
\\[-1.5ex]
2&$0.782004\cdots$&$\;\;\>0.53223180\cdots$&$   0.33355604(1)^* $&$   \;\;\>0.1988425(42) $&$0.1148728(43)$&$\;\;\>0.0649930(34)$&{$0.0362193(35)$}&$\;\;\>0.0199537(80)$\\
3&$0.625$&$\;\;\>0.2869495\cdots$&$   0.110252(1)^* $&$\;\;\>0.03888198(91)$&$0.01302354(91)$&$\;\;\>0.0041832(11)$&{$0.0013094(13)$}&$\;\;\>0.0004035(15)$\\
4&$0.506340\cdots$&$\;\;\>0.15184606\cdots$&$0.0357041(17)$&$   \;\;\>0.0077359(16)$&{$0.0014303(19)$}&$   \;\;\>0.0002888(18)$&$0.0000441(22)$&$   \;\;\>0.0000113(31)$\\
5&$0.414063\cdots$&$\;\;\>0.0759724807\cdots$&$0.0129551(13)$&$\;\;\>0.0009815(14)$&{$0.0004162(19)$}&$-0.0001120(20)$&$   0.0000747(26) $&$-0.0000492(48)$\\
6&$0.340941\cdots$&$\;\;\>0.03336314\cdots$&{$0.0075231(11) $}&$  -0.0017385(13) $&$   0.0013066(18) $&$-0.0008950(30)$&$   0.0006673(45) $&$  -0.000525(16) $\\
7&$0.282227\cdots$&$\;\;\>0.00986494662\cdots$&{$0.0070724(10) $}&$  -0.0035121(11) $&$   0.0025386(16) $&$  -0.0019937(28) $&$   0.0016869(41) $&$  -0.001514(14) $\\
8&$0.234614\cdots$&$-0.00255768\cdots$&{$0.00743092(93) $}&$  -0.0045164(11) $&$   0.0034149(15) $&$  -0.0028624(26) $&$   0.0025969(38) $&$  -0.002511(13) $\\[0.8ex]
\tableline
\end{tabular}
\normalsize
\vspace{0.5ex}
\caption[Numerical values of virial coefficients]{\centering Numerical
  values of virial coefficients. Values for $B_7$ $D>5$, $B_8$ $D>4$, $B_9$, and $B_{10}$
 are new, and other values improve on published
  literature results for $B_5$ and higher except for the results for
  $B_5$ for $D=2,3$ which are due to Kratky~\cite{kratky1982a}.} 
\label{tab:numericalvirial}
\end{sidewaystable}

\begin{table}[H]
\centering
\caption[Number of Mayer and Ree-Hoover diagrams]{\centering Number of Mayer and Ree-Hoover integrals}
\label{tab:diagram_number}
\vspace{2ex}
\small
\begin{tabular}{lccccccccc}
\tableline
\tableline
\\[-1.5ex]
& \multicolumn{9}{c}{Order}\\
& \multicolumn{1}{c}{2} & \multicolumn{1}{c}{3} & \multicolumn{1}{c}{4} & \multicolumn{1}{c}{5} & \multicolumn{1}{c}{6} & \multicolumn{1}{c}{7} &\multicolumn{1}{c}{8} & \multicolumn{1}{c}{9} & \multicolumn{1}{c}{10} \\[0.7ex]
\tableline
\\[-1.5ex]
Mayer&1 &1 &3 &10 &56 &468 &7123 & 194066 & 9743542\\
RH&1 &1 &2 &5 &23 &171 &2606 & 81564 & 4980756\\
RH/Mayer& 1& 1& 0.667& 0.500& 0.410& 0.365& 0.366 & 0.420& 0.511\\
RH, $D=1$ & 1& 1& 1& 1& 1& 1& 1& 1& 1\\
RH, $D=2$ & 1 &1 &2 &4 &15 &73& $\gtrsim$647 & $\gtrsim$8417& $\gtrsim$110529\\
RH, $D=3$ & 1 &1 &2 &5 &22 &161&$>$2334 & $>$60902& \\
RH, $D=4$ & 1 &1 &2 &5 &23 &169&$>$2556 & $>$76318 & \\
\tableline
\end{tabular}
\normalsize
\end{table}

\section{Virial Expansion}
\label{virial}

To calculate virial coefficients we use the reformulation of the Mayer
series~\cite{mayer1940a} carried out by Ree
and Hoover~\cite{ree1964c}. 

In the Mayer formulation virial coefficients $B_k$ are given as the sum of integrals which may be represented by biconnected graphs of $k$ points. We define a $k$ point graph $G$ as a set of vertices $V=\{v_i,i=1,\cdots,k\}$
 together with a set of edges $E$ whose elements are pairs $(u,v)$ where
 $u,v\epsilon V$. We will only be interested in undirected graphs, and not in
generalizations such as directed graphs or graphs with multiple edges
 and loops. A graph is biconnected if there are no vertices whose
 removal would result in a disconnected graph, which are more
 concisely known as articulation points. Then the virial coefficient
 at order $k$ is given in the Mayer formalism by 
\bea
B_k &=& \frac{1-k}{k!} \sum_{G\epsilon\mathscr{B}_k^L}S(G)  \nn
&=&\frac{1-k}{k!} \sum_{G\epsilon\mathscr{B}_k^U} C(G) S(G) \label{eq:bkmayer}
\eea
where $S(G)$ is the value of the integral represented by $G$,
 $\mathscr{B}_k^L$ is the set of all labeled, biconnected graphs,
 $\mathscr{B}_k^U$ is the set of all unlabeled biconnected
 graphs. $C(G)$ is a combinatorial factor which is defined as the
 total number of distinguishable labelings of a graph, otherwise given
 as: 
\be
C(G) = \frac{k!}{|{\rm Aut}(G)|}
\ee
for a graph with $k$ vertices, and $|{\rm Aut}(G)|$ is the cardinality of the automorphism group of the graph.

\begin{figure}[H]
\centering
\includegraphics[width=2.5cm]{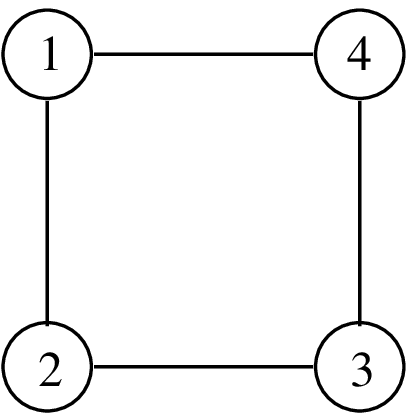}
\caption[Typical labeled graph]{\centering Typical labeled graph, $G_{\mathrm{ex}}$}
\label{fig:examplegraph}
\end{figure}

Each graph may be identified with an integral in which the vertices
represent coordinates in $D$--dimensional Euclidean space, and a bond
between vertices $i$ and $j$ represents the function \be f({\bf r_i
  -r_j})=\exp\left(-U({\bf r_i -r_j})/{k_BT}\right)-1 \ee For example,
the integral corresponding to the graph in \fig{fig:examplegraph} may
be written as 
\bea
S(G_{\mathrm{ex}}) &=& \lim_{V \rightarrow \infty} \frac{1}{V} \int d^D {\bf r_1} d^D
{\bf r_2} d^D {\bf r_3} d^D {\bf r_4} f({\bf r_{12}})
f({\bf r_{23}}) f({\bf r_{14}}) f({\bf r_{34}}) \nn 
&=&
\int d^D {\bf r_1} d^D {\bf r_2} d^D {\bf r_3} f({\bf r_{12}}) f({\bf
  r_{23}}) f({\bf r_{1}}) f({\bf r_{3}})
\label{eq:Mexamplegraph}
\eea
where ${\bf r_{ij}} = {\bf r_i - r_j}$, and $V$ is the volume to be
integrated over. In the second expression for the integral, ${\bf
  r_4}$ is defined as the origin. Naturally the value of the integral
does not depend on the labeling chosen, and this is the reason that in
\eqn{eq:bkmayer} the sum over labeled graphs can be converted to a sum
over unlabeled graphs.  

A useful
re-summation was performed by Ree and Hoover~\cite{ree1964a,ree1964c}
by introducing the function \be {\tilde f(\bf r)}=1+f({\bf
r})=\exp\left(-U({\bf r})/{k_BT}\right) \ee and then expanding each
Mayer graph by substituting $1=\tilde{f}-f$ for pairs of vertices not
connected by $f$ bonds. This method was previously used by Percus and
Yevick~\cite{percus1958a} in comparing the exact values of the fourth
and fifth 
virial coefficients with coefficients obtained from the
Percus--Yevick equation, and by Percus~\cite{percus1964a} in discussing
the derivation of the Percus--Yevick equation. One can see that upon
performing this re-summation on the Mayer series that vertices in the
resultant graphs will all be mutually connected by either $f$ bonds or
$\tilde{f}$ bonds, and that graphs must be biconnected with respect to
$f$ bonds. In this paper, we adopt the convention that graph edges
only correspond to $f$ bonds; in other contexts it can be very useful
to refer to the complement graphs in which the $\tilde{f}$ bonds are
edges. Now there is an additional factor for each labeled graph,
called the star content by Ree and Hoover, which may be either
positive or negative, and is equal to the total number of biconnected
subgraphs with an even number of edges removed (including zero),
subtract the number of biconnected subgraphs with an odd number of
edges removed. Thus the Ree-Hoover expression for the virial
coefficient $B_k$ is 
\bea
B_k &=& \frac{1-k}{k!} \sum_{G\epsilon\mathscr{B}_k^L}\mathrm{SC}(G)\tilde{S}(G)  \nn
&=&\frac{1-k}{k!} \sum_{G\epsilon\mathscr{B}_k^U} C(G) \mathrm{SC}(G) \tilde{S}(G) \label{eq:bkreehoover}
\eea
where $\mathrm{SC}(G)$ is the star content and $\tilde{S}(G)$ is the
Ree-Hoover integral value of the graph $G$. As a consequence of the
fact that for every $1$ expanded there is a positive and negative
coefficient, in the Ree-Hoover re-summation only the complete graph
which has no $1$'s contributes a net non-zero amount. Thus 
\be \sum_{G\epsilon\mathscr{B}_k^L}\mathrm{SC}(G) =  1 \label{eq:scsum}\ee

Now, to compare with the Mayer integral formalism of
\eqn{eq:Mexamplegraph} we give the expression for the Ree-Hoover
integral represented by \fig{fig:examplegraph} 
\bea
\tilde{S}(G_{\mathrm{ex}}) &=& \lim_{V \rightarrow \infty} \frac{1}{V} \int d^D {\bf r_1} d^D
{\bf r_2} d^D {\bf r_3} d^D {\bf r_4} f({\bf r_{12}})\tilde{f}({\bf r_{13}})
f({\bf r_{23}}) f({\bf r_{14}}) \tilde{f}({\bf r_{24}}) f({\bf r_{34}}) \nn 
&=&
\int d^D {\bf r_1} d^D {\bf r_2} d^D {\bf r_3} f({\bf r_{12}}) \tilde{f}({\bf r_{13}}) f({\bf r_{23}}) f({\bf r_{1}}) \tilde{f}({\bf r_{2}}) f({\bf r_{3}}) 
\label{eq:RHexamplegraph}
\eea

In the case of hard spheres, the potential is given by
\eqn{eq:hardspherepotential}, and so $f({\bf r})$ and $\tilde{f}({\bf r})$
are particularly
simple:
\bea f({\bf r}) &=& \left\{ \begin{array}{r} -1 \\ 0
\end{array} \right. \begin{array}{l} |{\bf r}|< \sigma \\ |{\bf
r}|>\sigma\\ \end{array}
\label{eq:fdef} \\ \tilde{f}({\bf r}) &=& \left\{ \begin{array}{r} 0 \\
+1 \end{array} \right. \begin{array}{l} |{\bf r}|<\sigma \\ |{\bf
r}|>\sigma \\
\end{array} \label{eq:ftildedef} \eea
The $f$ bonds force vertices to be close together, while $\tilde{f}$
bonds force vertices apart. These competing conditions may mean that
for some graphs contributing to $B_k$ in dimension $D<k-1$ there are
no point configurations that satisfy the conditions, and hence the
corresponding integral would be  zero for geometric reasons. 
Any configuration of points in $D$--dimensional Euclidean
space can contribute to at most one Ree-Hoover diagram, whereas it may
contribute to many Mayer diagrams, and in particular for all $k$ point graphs $G$
\be
|S(G)| \geq |\tilde{S}(G)|
\ee
The inequality is strict when one excludes the complete star
graph. This leads to the primary advantage of the Ree-Hoover
formulation over the original Mayer formulation:  it greatly reduces
the degree of cancellation between positive and negative diagrams. 

\section{Method}
\label{method}

In \sect{montea} we give a broad description for the Monte-Carlo
algorithm used to perform the numerical integration for our results of
\tab{tab:numericalvirial}. In \sect{notation} we define some graph
theory notation that is necessary for our detailed discussion of the
calculation of the star content of all unlabeled graphs of up to ten
points in \sect{star}, the determination of a minimal set of spanning
trees to generate configurations for the Monte-Carlo algorithm in
\sect{span}, and the unlabeling factor for graphs of up to ten points
in \sect{unlabel}. Lastly, in \sect{monteb} we give specific details
of the implementation of the algorithm. 

\subsection{Monte-Carlo Algorithm}
\label{montea}

We calculate virial coefficients via the hit or miss Monte-Carlo
integration algorithm described by Ree and
Hoover~\cite{ree1964a}. Monte-Carlo integration is appropriate for
high dimensional integrals, which is certainly the case for the
integrals we calculate as in general, the calculation of $B_k$ for $D$
dimensions is a $(k-1)D$ dimensional integral. 

If we need to integrate a function $F(x)$ over a complicated, high
dimensional domain $D_1$, the hit or miss Monte-Carlo method works by
enlarging the domain of integration to a simple, easily characterized
domain $D_2$ and setting the function to zero in $D_2 \setminus
D_1$. We then select points uniformly from within $D_2$ using a pseudo
random number generator, and the integral we are interested in is
given as the volume of the enlarged domain $V(D_2)$ multiplied by the
mean value of $F(x)$ over $D_2$. 
\bea
\int_{D_1} dx \; F(x) &=& \int_{D_2} dx \; F(x)  \hspace{1cm} D_1 \subset D_2, \; F(x) \equiv 0 \;\; \forall x \in D_2 \setminus D_1 \nn
&=&V(D_2) \lim_{N\rightarrow \infty} \frac{1}{N}\sum_{i=1}^N F(x_i) \hspace{1cm} x_i \in D_2
\eea 

The Ree-Hoover integral of a given $k$ point biconnected graph $G$ has
a trivial integrand which is $\pm 1$, but the domain is all
configurations $(r_1,r_2,\cdots,r_k)$ where $r_i\in \mathbb{R}^D$ and
the resulting point set may be identified with $G$. To calculate the
integral, we first identify a spanning tree of the graph, and use all
Ree-Hoover diagrams which are spanned by this tree as the enlarged
domain. To generate random configurations in the Monte-Carlo algorithm
we place the 
first particle at the origin, and then successively place the
succeeding particles randomly within unit balls according to the
generating spanning tree. Once we have the position of all particles
we check the inter-particle distances to determine if the distances
are greater than or less than one, and thus map the configuration to a
graph. The volume of the enlarged domain is straightforward to
calculate, as each edge of the spanning tree contributes a factor of
the volume of the unit ball in $D$ dimensions, which may also be given
in terms of $B_2$ as $(2B_2)^{k-1}$. 

The efficiency of the algorithm may be improved by taking in to
account all labeled isomorphs of $G$ which may be generated by the
spanning tree, but we then must divide by this unlabeling factor to
get the correct result. 

This method of calculating Ree-Hoover integrals may be extended to
calculating the virial coefficient itself, which up to a factor is
given by the sum of the Ree-Hoover integrals of all labeled
biconnected graphs multiplied by their star content. Thus we require a
set of spanning trees which are able to generate all biconnected
graphs, and we also need to calculate the unlabeling factor for each
of the biconnected graphs. Ree and Hoover~\cite{ree1964a,ree1964c},
Janse van Rensburg~\cite{jansevanrensburg1993a}, and Vlasov et
al.~\cite{vlasov2002a} proceed by finding a set of spanning trees, and
partitioning the biconnected graphs according to which spanning tree
will be used to calculate them. This means that if a given spanning
tree produces graphs which are not in its set then these
configurations are thrown away. We improve on this by combining the
unlabeling factors from each of the spanning trees as the sum of the
individual unlabeling factors, and never discard a geometric graph
that contributes to $B_k$. 

For each iteration of the Monte-Carlo calculation we use each of the
set of spanning trees to generate a configuration, which is mapped to
a graph. If the graph is biconnected then the contribution of this
graph, up to an overall factor, is given by the star content
multiplied by the number of possible labelings of the graph, divided
by the unlabeling factor. 

In practice, for each batch we keep a tally of all of the biconnected
graphs generated, and only condense this information using the star
content and unlabeling factor values to a value for $B_k$ at the end
of the loop. For this work we chose a batch size of $10^7$
configurations, and uncertainties were 
calculated using the standard formula 
\be
\mathrm{Error} = \left[\sum_{j=1}^q
\frac{(<I_j>-<I>)^2}{q(q-1)}\right]^{\frac{1}{2}}
\label{eq:montecarloerr}
\ee
where there are $q$ independent batches with value $I_j$. All
uncertainties given in this paper are equal to one standard deviation,
and the number of batches used in the calculation of the virial
coefficients in \tab{tab:numericalvirial} are given in
\tab{tab:batches} of Appendix \ref{spantable}.

\subsection{Graph Theory Notation}
\label{notation}

Here we define graph theory notation necessary for the description of
the calculation of the star content and unlabeling factor of
biconnected graphs, as well as the determination of a set of spanning
trees sufficient to generate all biconnected graphs of a particular
order. 

We will use the term subgraph of a graph G to describe graphs with the same
vertex set, but with an edge set that is a subset of the edge set of G. We will
allow G to be its own subgraph; when excluding this case we will refer
to proper subgraphs.

We define $\con_j(G)$ and $\bi_j(G)$ to be the set of subgraphs of $G$
with $j$ edges removed which are connected and biconnected
respectively. We define the composition of $\bi_i$ and $\bi_j$ by
applying $\bi_i$ to each element of the set $\bi_j(G)$ and taking the
union of the resulting sets. The end result is exactly
$\bi_{i+j}(G)$. Thus 
\be
\bi_i \circ \bi_j = \bi_{i+j}
\ee
However, if we apply $\bi_i$ to each of the $G^\prime \in \bi_j(G)$
then we can see that each subgraph in $\bi_{i+j}(G)$ is produced
exactly $\binom{i+j}{j}$ times. Thus 
\be
\sum_{G^{\prime}\in\bi_j(G)} |\bi_i(G^\prime)| = \binom{i+j}{j} |\bi_{i+j}(G)|
\ee
In particular, for $j=1$,
\be
|\bi_{i+1}(G)| = \frac{1}{i+1} \sum_{G^{\prime}\in\bi_1(G)} |\bi_i(G^\prime)|
\ee

We will define a minimally biconnected graph in analogy to the definition of
a tree: a minimally biconnected graph is a graph for which no proper subgraph
is biconnected. If we remove any edge from a minimally biconnected graph then
the resulting graph will not be biconnected.

An articulation star or clique cut-set of a graph is a proper subset
of vertices
that is maximally connected, i.e. edges connect each pair of vertices, and whose
removal will result in a disconnected graph.

Let CL be an algorithm that takes as input a graph G and defines
a relabeling of the vertices of
an input graph. Then we will call CL a canonical labeling algorithm
if it can be shown that
\bea
G_1^\prime &=& \mathrm{CL}(G_1)\\
G_2^\prime &=& \mathrm{CL}(G_2)
\eea
\be
G_1^\prime = G_2^\prime \Leftrightarrow G_1 \; \mathrm{and}\; G_2\; \mathrm{are \; isomorphic}
\ee

Thus, CL converts labeled graphs to unlabeled graphs, and this can, in
the appropriate context, greatly reduce 
the scale of a problem as there are many fewer unlabeled graphs by a
factor of approximately $k!$ at order $k$.

This is an NP-complete problem, and as such it is believed that it is
impossible to obtain an algorithm that can be guaranteed to determine
the canonical labeling of a graph in polynomial time in the number of
vertices $k$.

An efficient algorithm has been developed by
McKay\cite{mckay1981a} for the purpose of calculating the automorphism
group and also the canonical labeling of a graph. Through his own testing
on random graphs he finds performance for the canonical labeling
procedure is approximately O($k^\beta$) with $2<\beta<3$ for different
classes of graphs despite the fact that this is an
NP-complete problem. McKay's implementation of this algorithm in the 
C programming language, nauty
(no automorphisms, yes?), has been extensively used here.

\subsection{Star Content}
\label{star}

To calculate the star content of all biconnected nine and ten point graphs,
we use the formula of Ree and Hoover~\cite{ree1964c} outlined in
\sect{virial}, here given in the notation of \sect{notation}. 

\be \mathrm{SC}(G) = \sum_{i=0}^{|E|} (-1)^i | \bi_i(G) | \label{eq:list}\ee

We note that $| \bi_{0}(G) | = 1$ if $G$ is biconnected, and that 
$| \bi_i(G) | = 0$ for $i>0$ if and only if $G$ is minimally biconnected.
Hence we can enumerate all subgraphs with a certain number of edges
of a given biconnected graph, if we have this information for the subgraphs.
This naturally defines a recursive procedure, where if we apply this enumeration
algorithm to the complete star, and recursively apply it to all unlabeled
 subgraphs, we will generate this list and hence can calculate the star 
content.

It is more memory efficient, however, to enumerate all subgraphs for graphs with
$k$ edges (the Ree-Hoover ring), then to enumerate subgraphs of graphs with
 $k+1$ edges, until we finally reach the
complete star with $k(k-1)/2$ edges. As a by-product, all minimally biconnected graphs
up to tenth order were generated and recorded for use in the spanning tree
algorithm of \sect{span}.
Using this procedure the star content for all ten
point Ree-Hoover graphs was calculated in a matter of 30 minutes. We would
therefore expect that the star contents for $B_{11}$ may be calculated in
a matter of days, but the problem is significantly more demanding on
computer hardware as a large 
amount of memory is needed to perform the calculation efficiently.

We plot histograms of the number of 
labeled graphs versus star content
for $B_6$, $B_7$, $B_8$, $B_9$, and $B_{10}$, in  Figures 
\ref{graph:star6}--\ref{graph:star10} of Appendix
\ref{histograms}. For these figures we exclude graphs with zero star
content which have an articulation star, and it is for this reason
that the value appears to be low for $k=6,7$. For $k \ge 8$ the zero
star content value is far greater than the next largest value, and for
this reason is not shown in the plots. 
 Note that the histograms become sharply
peaked around zero star content, and that these plots become increasingly
symmetric as one increases the order.

It is natural that the star contents should be peaked around zero given the constraint imposed by \eqn{eq:scsum}. However, it is not obvious that there should be such a strong symmetry
between graphs with star contents of opposite sign, and it would be
desirable to find a reason for this behavior. 

\subsection{Spanning Trees}
\label{span}

To find a minimum or close to minimum set of spanning trees we
proceed as follows. We take the set of minimally biconnected graphs
obtained during the star content calculation procedure. 

We then need to enumerate all labeled spanning
trees of each of the minimally biconnected graphs, and use nauty to
help keep a list
of the unlabeled spanning trees of each graph. 

The problem of enumerating all spanning trees of a graph is considerably
more difficult. If a graph has a large number of spanning trees, then the
best available algorithms take time $O(N)$ with sub-leading additive
terms, where $N$ is the number of spanning
trees. There are many algorithms available which may be distinguished by
the sub leading terms, or in the amount of memory
required, and by whether they output each of the spanning trees, or
output the difference between consecutive trees.

We have implemented an algorithm~\cite{gabow1978a} which uses
$O(V+E+VN)$ time and $O(V+E)$ space. 
As we are only concerned with the case where $V$ is small, the resulting
increase in complexity does not matter much.

Naturally, each
biconnected graph must have at least one minimally biconnected graph
as a subgraph, and thus if we find a set of spanning trees that can
generate all minimally biconnected graphs then necessarily we will be
able to generate all biconnected graphs.

We could use all spanning trees as our set
without any problem. The reason we do not wish to do this in practice
is that some graphs such as the ring graph contribute significantly to
the final result, but only have perhaps one or two of a large number
of spanning trees that will be able to generate it.

We store this information in matrix form as follows:
$A_{ij} = 1$ then spanning tree $j$ is a subgraph of
minimally biconnected graph $i$, otherwise $A_{ij} = 0$. 
As posed, this is the minimum set cover
problem, for which
no known polynomial time algorithm exists. 

The minimum set cover problem may be defined as follows: given a
matrix in which all entries have the value of either 0 or 1, and there is at least one 1 in each row, then what are
the minimum number of columns that can be chosen such that there is
one or more 1's in each row? This problem has been shown to be NP-complete, and the authors could
not find implementations of algorithms that are efficient on average for
random matrices in the vein of the canonical labeling algorithm of McKay.

However, for our purposes we did not need the exact minimum, and we can apply the
greedy algorithm which will produce a minimal set cover (in the sense
that we cannot remove any spanning tree from the final result without
leaving a graph uncovered), that will be sufficiently close to the minimum
for our purposes. 
The greedy algorithm is defined
in the following way: choose the column with the
greatest number of 1s, where if there is a tie we break the tie arbitrarily, then
remove that column and the rows which have now been covered.
 Repeat until there are no rows remaining, at which point
the algorithm terminates and returns the numbers of the columns which
have been removed. We implemented a slight variation of this: at the
beginning of the main loop of the algorithm, if any row has a single 1
then the 
corresponding column is chosen initially, as it must be included
in any set cover solution.

We apply this algorithm to find a near minimum set of spanning
trees which are able to generate all biconnected graphs for orders
$k=5,\cdots,10$, and these are  
shown in \fig{fig:minspan} of Appendix \ref{spanfig}.

We may also optimize the procedure by choosing trees that will
minimize the error in the Monte-Carlo procedure by efficiently
calculating large diagrams. We could do this rigorously, by performing
the integration over a range of dimensions and orders, and obtaining a
different set of trees for each case. We note here that the
Ree-Hoover ring is one of the largest diagrams to the order we
calculate, and that the ring diagram and most other loosely packed
diagrams
 can be
efficiently generated by the tree without any leaves
(it is the only tree required to generate $B_5$ as can be seen in
\fig{fig:minspan}). For this reason 
we choose to have multiple copies of the tree without any leaves in
our set of spanning trees. 
We have compared this set with the set of all spanning trees
and it is more efficient, but we have not performed extensive testing
to justify this choice.

We make the comment that there is no difficulty in carrying this
 procedure out to much higher
order, as the number of trees and minimally biconnected graphs grow
relatively slowly with order compared with the rate of growth of the
 total number of graphs.

\subsection{Unlabeling Factor}
\label{unlabel}

Now that we have our minimal set of spanning trees, we need to
calculate the 
unlabeling factor as defined by Ree and Hoover.

The unlabeling factor is defined as the number of ways in which a
given spanning tree may generate a particular labeled
graph. Alternatively, we may define it as the number of labeled
isomorphs of a particular graph which have the (labeled) spanning
tree as a subgraph.

Ree and Hoover~\cite{ree1964a,ree1964c,ree1967a}, Janse van
Rensburg~\cite{jansevanrensburg1993a}, and Vlasov et
al~\cite{vlasov2002a} proceed by partitioning the set of biconnected
graphs in to those which will be calculated by spanning tree 1, 2 , 3,
etc., and determining the unlabeling factor for
each unlabeled graph. This is a computationally intensive procedure in
the same manner as for the star content calculation, although not
quite as intractable. This is because the number of labeled trees that
are subgraphs of the complete star is $k^{k-2}$. Thus we
will need to enumerate 
approximately $k^{k-2} 2^{k(k-1)/2}/k!$ spanning trees, which grows
very fast with order.

We improve on this method in two respects; firstly, we do not
partition the graphs, and allow the spanning trees to generate any of
the biconnected graphs. In this manner we do not throw out any
biconnected configurations that we generate. The combined effect of
unlabeling factors from all of the spanning trees is merely
the sum of the individual unlabeling factors. Secondly, in this formula 
we may use any unlabeled spanning tree more than once if we choose. As noted
in \sect{span}, we use a minimal set of spanning trees but with
repetitions of the tree with no leaves.

We then proceed as for the star content, as \eqn{eq:list} applies to
the operator for connected graphs $\con_j$, 
with the difference that we
need to keep track of all connected graphs (not just biconnected
graphs), and we only need to know the unlabeling factor of each graph
rather than the count of subgraphs partitioned by the number of edges
they have.

\subsection{Details of the Monte-Carlo Procedure}
\label{monteb}

All code is written in the C programming language, and the calculation
has been compiled and run on Linux machines with Pentium 4
processors. 

We use the random number generator of Ziff~\cite{ziff1998a}, and in
some cases we have confirmed results with the random number generator of
Knuth~\cite{knuth1997a}. 

We generate random points using the algorithm of
Banerjia and Dwyer~\cite{banerjia1993a} for $D=3,4,5$, and use similar
accept or reject algorithms  for higher
dimension. Alternative methods which do not require the rejection of
any generated points, such as using Gaussian variables, may be faster
for higher dimensions, but as this is not the bottleneck step in the
algorithm we did not implement or test such alternatives. 
 
Given a set of $N_{\mathrm{span}}$ spanning trees which will be used
to generate configurations, we wish to be able to generate
configurations as fast as possible by minimizing the number of random
points to generate. To generate each tree we will need to have $k-1$
random points and hence naively we need a total of
$(k-1)N_{\mathrm{span}}$ random points for each iteration. 
By using the same set of random points for each spanning
tree we only need to generate $k-1$ random points per iteration.

The remaining detail of the Monte-Carlo calculation that needs to be
discussed is how to identify graphs as they are being generated. The
method used was to create a hash table of all canonically 
labeled biconnected graphs with non-zero star content, so that when a graph is
generated one can calculate the canonical label and check if it is in
the hash table using linear probing~\cite{knuth1997a}. In practice,
the canonical labeling 
algorithm of McKay~\cite{mckay1981a} is quite slow  compared to the
speed at which configurations 
are generated, and so if we possible we wish to avoid this step.

For small graphs, which for the computers on which this program was
run means for eight or less vertices, we can create an array with
$2^{k(k-1)/2}$ elements for which the address is identified with all
labeled graphs of $k$ points. One can map any graph to a sequence of
$k(k-1)/2$ bits by imposing an order on all vertex pairs, and then
giving the bit corresponding to vertices $i$ and $j$ a value of 1 if
$(i,j) \in E$, and 0 otherwise. Thus once a graph is generated the
canonical label can be quickly looked up in this array provided some
pre-processing is done before the main loop of the Monte-Carlo
integration. Unfortunately, as the total number of labeled graphs
grows extremely rapidly with order, memory limitations mean that even
on a supercomputer it would not be possible to do this for graphs with
more than 9 vertices. 

For large graphs it is therefore much more difficult to perform the
identification quickly, and so we attempt to identify if the graph has
non-zero 
star content prior to calculating the canonical label. In particular
contributing graphs must be biconnected, and cannot contain an articulation
star, as discussed by Ree and Hoover~\cite{ree1964c}. To this end,
a graph that has been generated
is first checked
to see if there are any vertices of degree 1, and then tested for
biconnectivity, and then checked to see if it has an articulation 
star using an algorithm described below.
If any of these
conditions are met, then the star content is necessarily zero (although
many graphs with zero star content do not have an articulation star);
otherwise 
the canonical label is calculated using the canonical labeling 
algorithm of McKay~\cite{mckay1981a}.

Ree and Hoover~\cite{ree1964a} showed that graphs with an articulation
star will have zero star content, although the converse is not true:
some graphs with zero star content have no articulation star for
orders greater than 5. An articulation star, otherwise known as a
clique separator in the mathematical literature, is a completely
connected subset of vertices of a graph whose removal disconnects the
graph. A biconnected graph is a graph with no articulation points, and
so the set of graphs with no articulation star is a natural
generalization. 
For the purpose of identifying graphs with an articulation star,
 we implement an
algorithm~\cite{rose1976a,tarjan1984a,tarjan1985a} which finds all
articulation stars by finding a minimum ordering
of the graph, and calculating the resulting fill-in graph. In
practice, we used an algorithm which finds minimal rather than minimum
orderings as this much faster; the trade off is that some graphs with
an articulation star will not be identified, but for graphs generated
in the calculation of $B_9$ and $B_{10}$ this occurred only about 2
percent of the time. Note that a graph is never falsely identified as
having an articulation star when it does not. 

An alternative method would be to implement the
algorithm of Whitesides~\cite{whitesides1981a} in which a single
articulation star is found in the same asymptotic time as all
articulation stars are found by the minimal ordering method. However,
the method of Whitesides perhaps should be explored as we are
interested in small graphs, where the asymptotic behavior of the
algorithm is perhaps not as important as the overhead involved in the
calculation. 

Given the star content, and unlabeling factor, we may then proceed to
calculate the virial coefficients up to order ten, in any integer
dimension. The results of these computations have been given in 
\tab{tab:numericalvirial}.

\section{Background}
\label{background}

We describe in this section the phase transition undergone by hard
spheres in Subsection \ref{phasetransition}, and the rigorous
information we have about this system in Subsection
\ref{rigorous}. Ideally, we would like to study the relation of the
phase transitions to the virial coefficients.
However the unambiguous  interpretation of the ratio plots 
and their relation to the
freezing transitions seen in computer studies is
made difficult by the fact that we have very few exact results
available for the hard sphere system beyond the general theorems that
the pressure is continuous and non-increasing for positive densities in
the physical region. For this reason in Subsection \ref{scenarios} we
discuss the various scenarios that have been proposed for the position
of the leading singularity of hard spheres in order to provide a
framework for later discussion in Section \ref{discussion}.

\subsection{Phase Transition}
\label{phasetransition}

Perhaps the property of hard spheres which is the most interesting and
surprising characteristic is the existence of a phase
transition, as discovered for $D=3$ by Alder and
Wainright~\cite{alder1957a} and Wood and Jacobson~\cite{wood1957a} through
molecular dynamics simulations in 1957, and for $D=2$ in
1962~\cite{alder1962a}. For
dimensions three, four and
five~\cite{michels1984a} the phase transition is believed to be first
order, while for 
hard discs the situation is more controversial despite intense research efforts over the past forty 
years~\cite{alder1962a,jaster1999a,dash1999a,binder2002a}. Recently
Jaster~\cite{jaster2004a} showed that as the density of the system
increases that hard discs first undergo
a first or second order transition from the fluid phase to a hexatic phase with short range
positional and quasi long range
orientational order, and then undergo another second order transition
to the solid phase which has quasi long range positional and
orientational order. This is consistent with the 
Kosterlitz--Thouless--Halperin--Nelson--Young scenario, for which both
transitions must be second order.

The phase
transition freezing ($\rho_f$) and melting ($\rho_s$) 
  densities are given in \tab{tab:rho} for
  $D=2$~\cite{jaster1999a,jaster2004a},
  $D=3$~\cite{alder1957a,alder1960a,hoover1968a}, and
  $D=4,5$~\cite{michels1984a}. We list also
the density $\rho_{cp}$ and the scaled density
$B_2\rho_{cp}=2^{D-1}\eta_{cp}$ of the 
densest lattices for dimensions 
$D=2,\ldots,8$ from the lattice catalogue of Nebe and
Sloane~\cite{sloaneweb}. Note that the density $\eta=1$
corresponds to all space being filled.
Finally, we include the lower
bound of the radius of convergence obtained by Lebowitz and
Penrose~\cite{lebowitz1964a}.

\begin{table}[H]
\centering
\caption[Values for $B_2$ and the density of the closest packed
  lattices for $D=2,\cdots,8$. ]{Values for $B_2$, the density of
  the closest packed lattices, the densities at which freezing
  ($\rho_f$) and melting ($\rho_s$) occur, and the bound of Lebowitz and
Penrose~\cite{lebowitz1964a}, for hard spheres of diameter $\sigma$ in
  dimensions $D=2,\cdots,8$.}
\label{tab:rho}
\vspace{2ex}
\begin{tabular}{rrcrrrr}
\tableline
\tableline
\\[-1.5ex]
\multicolumn{1}{c}{$D$} & \multicolumn{1}{c}{$B_2$} &
\multicolumn{1}{c}{$\rho_{cp}$}& 
\multicolumn{1}{c}{$B_2\rho_{cp}=2^{D-1}\eta_{cp}$}&
\multicolumn{1}{c}{$\rho_f/\rho_{cp}$}&\multicolumn{1}{c}{$\rho_s/\rho_{cp}$}
&\multicolumn{1}{c}{$\rho_{LP}/\rho_{cp}$}\\
\tableline
\\[-1.5ex]
2  & $\frac{\pi\sigma^2}{2}$ & $\frac{2}{\sqrt{3}\sigma^2}$ &
$1.8137\cdots$ &$0.78$&$0.81$&$0.03990$\\[0.3ex]
3  & $\frac{2\pi\sigma^3}{3}$ & $\frac{\sqrt{2}}{\sigma^3}$ &$2.9619\cdots$&$0.66$&$0.75$&$0.02444$\\[0.3ex]
4  & $\frac{\pi^2\sigma^4}{4}$ & $\frac{2}{\sigma^4}$ &$4.9348\cdots$&$0.50$
&$0.68$&$0.01467$\\[0.3ex]
5  & $\frac{4\pi^2\sigma^5}{15}$ & $\frac{2\sqrt{2}}{\sigma^5}$ &$7.4441\cdots$&$0.41$&$0.62$&$0.00971$\\[0.3ex]
6  & $\frac{\pi^3\sigma^6}{12}$ & $\frac{8}{\sqrt{3}\sigma^6}$ &$11.9343\cdots$&&&$0.00605$\\[0.3ex]
7  & $\frac{8\pi^3\sigma^7}{105}$ & $\frac{8}{\sigma^7}$ &$18.8990\cdots$&&&$0.00382$\\[0.3ex]
8  & $\frac{\pi^4\sigma^8}{48}$ & $\frac{16}{\sigma^8}$ &$32.4696\cdots$&&&$0.00224$\\[0.8ex]
\tableline
\end{tabular}
\normalsize
\end{table}

The existence of singularities foe $D\geq 3$ at $\rho_f$ and $\rho_s$
is much more controversial than the singularities for $D=2$. 
It has long been argued by Fisher~\cite{fisher1965b} that at
these phase boundaries the pressure will be infinitely differentiable
but will not be analytic and thus cannot be analytically continued
into the region $\rho_f<\rho <\rho_s$ from either side. These
singularities have been proven to exist in the Ising model by
Isakov~\cite{isakov1984a}, but for hard spheres nothing rigorous has
been proven. 
In this scenario the freezing density $\rho_f$ can be determined in
principle from the leading singularity on the positive real axis of
the low density equation of state.

The alternative view of the first order (freezing) transition in hard
spheres for $D \geq 3$ is the assumption that there are no
singularities at the phase boundaries and that analytic continuation
from both sides is possible.
The freezing transition
is seen as a Maxwell construction making a convex envelope from a low
and a separate high density free energy. In this scenario it is
impossible to determine the freezing density from the low density of
state alone.

It is obviously of  great theoretical importance to determine which of
these two scenarios is correct for hard sphere for $D\geq 3.$ This
is a particularly difficult question if the radius of convergence
of the virial series is determined by a singularity in the complex
plane, which the ratio plots indicate is very likely the case for
$D\geq 4$. In this case
if the radius of convergence is less than the freezing density
$\rho_f$ it is necessary to find a way to analytically continue the
virial expansion beyond the radius of convergence to study a possible
singularity at $\rho_f.$ It is surely not possible to extract from our
10 term series both a leading singularity in the complex plane and to
reliably continue the expansion beyond the radius of convergence to
detect an infinitely differentiable singularity at $\rho_f.$
Consequently we will restrict our attention to locating the
leading singularity in the complex plane.  

\subsection{Rigorous Results}
\label{rigorous}

There are few rigorous results available for the problem of hard
spheres, and we will briefly summarize those few results here.

Groeneveld~\cite{groeneveld1962a} proved that for the cluster
expansion the radius of convergence
must be greater than $1/(2eB_2)$, where $e=2.71828\cdots$, and less
than $1/(2B_2)$. In addition, it is known that the cluster
coefficients alternate in sign and hence the leading singularity is on
the negative, real fugacity axis. Lebowitz
and Penrose~\cite{lebowitz1964a} 
adapted this bound to derive a corresponding expression for the virial
series, given in \eqn{eq:lebowitz}. 
\bea
\rho_{LP} &\ge& \frac{0.14476}{2B_2}\nn
\eta_{LP} &\ge& \frac{0.14476}{2^{D}}
\label{eq:lebowitz}
\eea
Thus we know that the pressure is
an analytic function of 
density for small densities; unfortunately in practice this bound does
not seem to be close to the true radius of convergences as it is a
long way from the density at which the phase transition
is seen to occur, as can be seen from \tab{tab:rho}.

Later Fisher~\cite{fisher1965a} obtained bounds on the pressure in the
vicinity of close packing for a $D$--dimensional system of hard
particles. Extending the results of Hoover~\cite{hoover1965a} for hard
parallel cubes, Fisher was able to prove that for
$\Upsilon \equiv \frac{\rho_{cp}}{\rho}-1 \rightarrow 0$ that
\be
\frac{\xi_1 D}{\Upsilon} \leq \frac{P}{\rho k_B T} \leq \frac{\xi_2
  D}{\Upsilon}
\ee
where $0<\xi_1<1<\xi_2<\infty$.
For the general case of hard core particles, which includes hard
spheres, he was able to establish the weaker relation
\be
\eta_1\ln\Upsilon^{-1} \leq \frac{P}{\rho k_B T} \leq
\frac{\eta_2\ln\Upsilon^{-1}}{\Upsilon}
\label{eq:freevolume}
\ee
for any $ \eta_1 < 1 < \eta_2$.

\subsection{Scenarios for the Position of the Leading Singularity}
\label{scenarios}

We present here various scenarios that exist for the location and
nature of the leading singularity of the virial series for hard
spheres in $D$ dimensions.

Over the past 40 years many approximate equations of state have been
proposed for hard spheres, most commonly for the case $D=3$. These
approximates may be categorized by the location of their leading
singularity. 

Many have high order poles at the space filling density $\eta=1$,
including the solutions of the compressibility and virial
Percus--Yevick integral 
equations~\cite{percus1958a} for
$D=3$ by Thiele~\cite{thiele1963a} and
Wertheim~\cite{wertheim1963a,wertheim1964a}, 
the scaled particle theory of Reiss, Frisch, and
Lebowitz~\cite{reiss1959a}, a proposal by
Guggenheim~\cite{guggenheim1965a}, and the widely used empirical
formula of Carnahan and Starling~\cite{carnahan1969a}.

The equations of state of Goldman and White~\cite{goldman1988a} and
Hoste and Dael~\cite{hoste1984a} have simple poles at or near the
packing fraction of closest
packed hard spheres. 

Other equations of state have a fractional power law divergence at or
near the ``random close packed'' density  $\eta_{{rcp}}=0.64$ as
defined
by~\cite{bernal1960a,bernal1964a,scott1960a,finney1970a}. These
approximates are obtained by constraining the divergence of the
leading singularity to be of the form
\be
Pv/k_BT=A(\eta-\eta_{{rcp}})^{-s}
\ee
As an
example $s$ is estimated as 1 in~\cite{lefevre1972a} as $0.678$
in~\cite{ma1986a} 
and $0.76$ in~\cite{song1988a}. In~\cite{jasty1987a} other values 
of $\eta_{{rcp}}$ are chosen and the values of $s$ lie in 
the range $0.6\leq s\leq 0.9$
depending on the approximation used.

Some approximate equations of state include the freezing density, and
we mention in particular that Torquato
\cite{torquato1995a,torquato1995b} proposes an equation of state which
agrees with the equation of Carnahan and Starling~\cite{carnahan1969a}
for $\eta < \eta_f$ but which is of a
different form for $\eta > \eta_f$. 

For dimensions greater than 3, we note the analytic solution by Leutheusser~\cite{leuthesser1984a}
of the Percus--Yevick equation in $D=5$, and the recent work of Robles, L{\'o}pez~de~Haro,
and Santos~\cite{robles2004a}, in which the analytic solution was obtained for the Percus--Yevick equation in $D=7$. One of the most interesting results is
that the radius of convergence is no longer determined by the
singularity on the positive real axis at
$\eta=1$, but instead by a singularity on the negative real axis for
$D \ge 5$.

Baram and Luban~\cite{baram1979a} in 1979 fitted the first seven virial
coefficients using 
Levin approximants, and
concluded that the leading singularity is at the close packed density.

More recently, in 1994 Sanchez~\cite{sanchez1994a} fitted the virial
series with a Pad\'e 
approximant, and then expressed the density in terms of the
compression factor by inverting the series. He then fitted the inverted
series with Pad\'e approximants of increasing order to obtain
estimates of the density at which the compression factor diverges. Sanchez
then argued that as the order of the Pad\'e approximants was increased
that the position of the leading singularity converged to the
density of close 
packing in two and three dimensions, and this was taken as evidence that
the leading singularity of the virial series is in fact the density of
closest packing.

Gaunt and Joyce~\cite{gaunt1980a} provided a counter argument to the
conclusion and methods of Baram and Luban, and indeed to Sanchez
despite the fact that the work of Sanchez came well after. They
performed a ratio analysis~\cite{guttmann1989a} of the virial
series, which they advocated as a robust method in the absence of any
knowledge of the exact form of the function being described by the
series. What was then the seven term virial series appeared to be
behaving smoothly and ratio analysis suggested that the leading
singularity lies on the positive real
axis, perhaps at the space filling density. However,
they pointed out that there is good reason to be cautious in extracting
asymptotic behavior from only the low order coefficients: for several
varieties of
lattice gas with either nearest neighbor exclusion or first and second
nearest neighbor exclusion, they show that the ratios initially behave
smoothly indicating a singularity on the positive real axis, and then
suddenly change behavior, and eventually change sign. This is because
the asymptotic 
behavior is determined by a complex conjugate pair of singularities
that mask the physical singularity on the real axis.

The only models which demonstrate a phase transition for which
exact results are available are hard core lattice gases in two
dimensions. In particular the hard hexagon
model~\cite{baxter1980a,richey1987a,joyce1988a} and a model of hard
squares~\cite{baxter1970a} have been exactly solved, and in both cases
the radius of
convergence is limited by a complex conjugate pair of singularities thus
resulting in virial coefficients that oscillate in sign. 

Other relevant models include that of hard parallel
cubes~\cite{hoover1962a} for which negative 
virial coefficients have been seen as low as sixth order, along with
the Gaussian model~\cite{uhlenbeck1962a,baram1991a,clisby2004c}
for which negative
coefficients can be seen for dimensions $D \ge 1$. In particular for
the Gaussian model oscillations are seen in the sign of the virial
coefficients and this is evidence that the dominant singularities to
are in the complex plane.

Virial coefficients have been calculated for hard
ellipsoids
and other hard particles in two and three dimensions~\cite{vlasov2002a}.
 For these models
we also see negative coefficients at relatively low order, with the
effect being more dramatic for shapes which are far from spherically
symmetric; in the case of ellipsoids, the effect is dramatic for large
aspect ratio.

The phase transition behavior of hard core systems ultimately depends
on the geometry of the crystalline phase. Thus it is not immediately
obvious how relevant hard hexagons and hard parallel cubes are to the hard
sphere problem, as the crystalline phase fills all space and the
phase transition has been observed to be second rather than first
order. However, one
could argue that this is only important in the vicinity of the phase
transition itself, and that the leading singularities of the virial
series for $D \ge 2$ may be a conjugate pair
of singularities in the complex plane.

\section{Discussion}

\label{discussion}
The most important property of the virial coefficients $B_k$ is  not
their actual numerical values for $k$ less that some finite number but
rather their asymptotic behavior as $k\rightarrow \infty$ because it
is the asymptotic value which determines the radius of convergence. Of
course no finite number of virial coefficients can give information on
the $k\rightarrow \infty$ behavior unless there is some {\it a priori} 
reason to expect that the values of $k$ are already in the asymptotic
$k\rightarrow \infty$ regime.

In Subsection \ref{zero}
 we propose two such criteria for the asymptotic regime of hard
spheres and will see that for $k\leq 10$ there is no dimension in which
both criteria are fully satisfied. Nevertheless it is still of
interest to determine what results are obtained if well known methods
are used in an attempt to determine the asymptotic behavior of the
 series from the first ten virial 
coefficients. In Subsections \ref{ratio} and \ref{diff} 
 we proceed with the methods of ratio analysis and
 of differential approximants respectively as
 robust and general methods for the purpose of series
 analysis. 

\subsection{Criteria for asymptotic behavior of $B_k$}
\label{zero}

Classical systems of hard particles for which the interaction
potential 
is either infinite or zero, such as hard spheres and parallel 
hard cubes, are special in that some graphs have Ree-Hoover 
integrals which are identically zero for dimensions $D$ less than 
some maximum value, as the restrictions imposed by the graph are so 
many that no configuration of points may satisfy them. Hence the 
number of Ree-Hoover integrals contributing at a particular order 
depends on dimension.
We study
this problem by performing the Monte-Carlo calculation outlined previously,
but in addition we keep track of how many diagrams have been 
generated at least once from batch to batch whereas previously 
this information was thrown away.

We tabulate our results for the number of contributing Ree-Hoover 
integrals in \tab{tab:diagram_number}. For $D=1$,
only the complete star graph is non-zero, for $D=2$ many graphs are zero at the
orders calculated, but for $D>2$ it is difficult to make any estimate
of how many zero diagrams there are. This is because there are a large number
of graphs with numerically small Ree-Hoover integrals, and there is 
no way to estimate what the distribution of the values
of these small diagrams is. Without this information it is 
impossible to estimate the number of zero
diagrams until the Monte-Carlo procedure has run for a 
sufficiently long time such that
almost all non-zero diagrams have been generated. 
It was found to be impossible to
reach this regime for $B_9$ and $B_{10}$ in dimensions $D=3,4$, 
and hence we suggest that new
direct ways must be found to count zero diagrams if progress 
is to made on this problem
for dimensions greater than two. 

The rate of growth of the number of non-zero Ree-Hoover 
integrals in two dimensions is far less than that of the number of 
biconnected graphs with non-zero star content. It is possible that 
the rate of growth has been reduced from $2^{k(k-1)/2}$ to something 
like exponential growth, but there is not enough data to make 
a definitive statement on this issue.

\vspace{.1in}

{\bf Criterion 1}

The number of nonzero Ree Hoover diagrams has approached its large $k$
behavior.

\vspace{.1in}

For $k=10$ this criterion is only completely fulfilled for $D=2$ and is not
fulfilled at all for $D\geq 5.$

Our second criterion has been presented in our previous paper
\cite{clisby2004b}

\vspace{.1in}
{\bf Criterion 2}

The loose packed diagrams with the number of $\tilde f$ bonds near
their maximum value numerically dominate $B_k.$

\vspace{.1in}

The validity of this criterion has been studied in detail in
\cite{clisby2004b}. Here it was seen that for $D=3$ and $k\geq 12$ the
criterion is well satisfied and that as $D$ increases the criterion is
satisfied for smaller values of $k$. However, for $D=2$ the criterion
was not satisfied even for $k$ as large as 18.

We thus conclude that there is no dimension in which both of these
criteria are fully fulfilled although for $D=3$ and $D=4$ it is possible
that they both could hold for some moderate values of $k$ of the order
of 12 to 14.

\subsection{Ratio Analysis}
\label{ratio}

We begin our analysis of the virial coefficients given in
 \tab{tab:numericalvirial}  by
making a ratio analysis~\cite{gaunt1980a,guttmann1989a}. 
By comparison with the test function
\begin{equation}
\sum_{k=1}^{\infty}a_kz^k~~~{\rm with}~~a_k=k^s/z^k_c,
\end{equation}
which as $z\rightarrow z_c$ has the singularity
\begin{equation}
(1-z/z_c)^{-1-s},
\end{equation}
we see, up to possible logarithms which cannot possibly be seen in
our ten term series, that the $k\rightarrow \infty$ behavior of the 
ratios  
\begin{equation}
a_{k+1}/a_k\sim z_c^{-1}(1+s/n+O(n^{-2}))
\end{equation}
indicates a simple pole at $z=z_c$ if $s=0$ and a divergence more
(less) singular than a pole if $s$ is positive (negative). Note in
particular that if $s<-1$ that the function is finite at $z=z_c$ 
and if $s<-1-n$ the
first $n$ derivatives are also finite at $z=z_c$ even though there is
a divergence in the $(n+1)^{st}$ derivative. 

We normalize the ratios to the density of the closest packed
lattice by plotting $B_k\rho_{cp}/B_{k-1}$ and these ratios are
plotted versus $1/k$ in Figures \ref{graph:ratio1}--\ref{graph:ratio5}
for $D=2,\cdots,8.$  
The values of $\rho_{cp}$
obtained from the catalogue of lattice packings of Nebe and Sloane
\cite{sloaneweb} are given in \tab{tab:rho}. 

The qualitative description of our results is as follows.

\subsubsection{$D=2$}

The plot of $B_k\rho_{cp}/B_{k-1}$ for $D=2$ is given in \fig{graph:ratio1}. The
ratios are all positive and are smoothly decreasing. At $k=10$ 
the ratio is $0.9992$ which is just
less than the value of unity it would have if the leading singularity were at
close packing. However, the ratios are still falling and hence suggest
the radius of convergence is determined by a singularity at a density
greater than close packing. 
For comparison the ratios for $D=3$ are also plotted.

\subsubsection{$D\geq 5$}

The ratios for $D\geq 5$ are plotted in Figures \ref{graph:ratio2} and \ref{graph:ratio3}. In these cases
the first few virial coefficients are positive (giving positive
ratios) and then to the order given the coefficients alternate in sign
(giving negative ratios). This suggests that there is a singularity on
the positive real axis which dominates the series initially, but 
there is a second singularity  in the complex plane or negative real
axis which is closer to the origin which determines the actual radius
of convergence and dominates the ratios at high order.     
If the leading singularity is on the negative real axis the
ratios will smoothly converge to some negative value, otherwise the
ratios will oscillate. 

\subsubsection{$D=4$}

The ratios for $D=4$ are plotted in \fig{graph:ratio4}. There are no negative
virial coefficients for $k\leq 10$ but even though the accuracy for
$B_{10}$ is limited there seems to be a very definite oscillation which
is developing just as oscillations develop for $D\geq 5.$ Hence, from
examining \fig{graph:ratio4} we expect that negative virial coefficients will
occur for $k$ not much greater than 12.

\subsubsection{$D=3$}

The most important case is $D=3$ and this is also the most difficult
to interpret. This case is plotted in \fig{graph:ratio2}  where it is compared with 
with $D=2$ and is plotted on an expanded scale in \fig{graph:ratio5}. In the plot
of \fig{graph:ratio2} the ratios for $D=3$ appear much the same as the ratios for $D=2$
and extrapolate to a density significantly larger than
$\rho_{cp}$. However, an inspection of \fig{graph:ratio5} reveals that  unlike
the situation for $D=2$ the ratio plots for $D=3$ are not always
convex and can be seen from the magnitude of the slopes of the line
segments given by 
$5.8959\cdots,4.43686(21),2.80402(85),2.212(32),2.279(15),1.745(75),$
and $1.34(34)$. This lack of monotonicity is evidence that the ratios
for $D=3$ are 
displaying small oscillations in amplitude. If these oscillations grow
in magnitude then may be expected that eventually there will be
oscillations in the signs of the virial coefficients and the radius
if convergence will be determined by a singularity away from the positive
real axis.

\subsection{Pad\'e and Differential Approximants}
\label{diff}

\subsubsection{Notation}

The simplest way to make extrapolations of a finite number of terms in
a power series expansion is the Pad{\'e} approximant which assumes
that the power series represents a rational function f(z) expressed as
\begin{equation}
f(z)=\frac{P_M(z)}{Q_N(z)}
\end{equation}
where $P_M(z)$ and $Q_N(z)$ are polynomials of degree $M$ and $N$
respectively. We refer to this approximant as $[M/N].$ The
coefficients of these polynomials are easily computed from the power
series of $M+N-1$ terms by solving a system of linear equations
This form cannot possibly represent a pressure which is continuous at
its leading singularity and thus is not appropriate for investigating
the question of equations singularities in equations of state for
the fluid phase of any system. Nevertheless Pad{\'e} approximates have
been used by many
authors~\cite{ree1964a,ree1967a,jansevanrensburg1993a,sanchez1994a}
either to obtain information about the singularities of the virial
series or as approximate
equations of state for hard spheres. For reasons of comparison with
the literature we list several Pad{\'e} approximants for $D=2,3$ in
the text below. 

A much more general method for extracting information from power series are the
differential approximants presented in
detail by Guttmann~\cite{guttmann1989a}. In this method the power series is
assumed to be represented by a function  $f(z)$ which is the 
solution of the linear differential equation
\begin{equation}
\sum_{i=0}^K Q_i(z;L_i)z^i\frac{d^i}{dz^i} f(z)=R(z;M).
\label{eq:diff}
\end{equation}
where $Q_i(z;L_i)$ and $R(z;M)$ are polynomials of degrees $L_i$ and
$M$ respectively
\begin{equation}
Q_i(z;L_i)=\sum_{k=0}^{L_i}q_{k;i}z^k,~~~R(z;M)=\sum_{k=0}^M r_kz^k
\end{equation}
This form will incorporate the feature that
$P(\rho)/k_BT=\rho+O(\rho^2)$ for  $z\rightarrow 0$ by requiring that
\begin{equation}
r_0=0,~~~{\rm and}~~~ r_1=q_{0,1}+q_{0,0}
\end{equation}
The form (\ref{eq:diff}) is sufficiently general that it allows algebraic
(or possibly logarithmic) singularities
 at the $L_K$ zeroes $z_j$ of $Q_K(z;L_K)$ The behavior near 
these zeroes $f(z)$ in the case where the zeros are simple is
\begin{equation}
f(z)\sim A(z_i)|z-z_i|^{\alpha_i}+{\frac{R(z_i;M)}{Q_0(z_i;L_0)}}
\end{equation}
with
\begin{equation}
\alpha_i=K-1-\frac{Q_{K-1}(z_i;L_{K-1})}{z_iQ'_K(z_i;L_K)}
 \end{equation}

The special case $K=1$ and $R(z,M)=0$ is called a Dlog Pad{\'e}
approximant and in this case $f(z)$ is explicitly given as
\begin{equation}
f(z)=z \exp \left\{-\int_{0}^{z}dz \frac{Q_0(z;L_0)+Q_1(z;L_1)}{
  zQ_1(z;L_1)}\right\}
\end{equation}
which near the zeroes $z_i$ of $Q_1(z;L_1)$ behaves as
\begin{equation}
f(z)\sim A(z_i)|z-z_j|^{-\frac{Q_0(z_j;L_0)}{ z_iQ'_1(z_i;L_1)}}.
\end{equation}
This $f(z)$ either diverges or vanishes at the singular points $z_i$
and thus, like the Pad{\'e} approximant the Dlog Pad{\'e} 
is not able to accurately represent a singularity in 
the pressure at a point such as the
freezing transition where the pressure is continuous. 
However, it should be adequate to study the question of whether for
$D=3$ the leading singularity is in the complex plane off the
positive real axis.

\subsubsection{Method}
\label{diffmethod}

We fitted the series
with regular Pad\'e, Dlog Pad\'e and differential approximants,
and attempted 
to get some measure of how well these approximants represented the
series by seeing how effective they were at ``predicting''
coefficients when given a shortened series. This was done
systematically by evaluating a measure which would quantify the 
relative mean square deviation of predicted coefficients from the 
actual coefficients of \tab{tab:numericalvirial}. We reached the
conclusion that there was no difference between Pad\'e, Dlog Pad\'e
and differential approximants in extrapolating the virial series for
hard spheres given the coefficients $B_1, \cdots , B_9$.

The Fortran program NEWGRQD given by
Guttmann~\cite{guttmann1989a} was used
to determine the differential approximants. 
We will now discuss the information that can be drawn from the
approximants in each dimension in terms of the leading singularities,
expected asymptotic behavior, and predicted coefficients. 

The scatter in the position and exponent of
singularities, and in predicted coefficients comes from two sources:
uncertainties in the virial coefficients, and differences between the
approximants themselves. We take in to account the virial coefficient
uncertainties by generating a reasonably large set (1000) of series by
sampling from Gaussian distributions with the center given by the best values
listed in \tab{tab:numericalvirial}, and the width given by the
corresponding uncertainties. We then determine the Pad\'e and
differential approximants corresponding to each of these series, and
combine the results to get mean values with uncertainties for
the predicted coefficients of each approximant. In
addition we obtain uncertainties for the coefficients of the rational
function $P_L(\rho)/Q_M(\rho)$ for the Pad\'e approximants. We found
that the scatter in the predicted series coefficients due to
uncertainties for Pad\'e approximants matched the scatter among
different approximants quite 
closely. The differential approximants were considerably more
sensitive to the uncertainty of the coefficients, and generally were
stable for around 2 to 3 less coefficients than the Pad\'e
approximants. Below we will neglect the discussion of scatter due to
uncertainties in the coefficients and only consider the scatter
between approximants. 

We list the 
singularities and corresponding exponents for the differential approximants for $D=2,3$ in
Tables \ref{tab:diff2} and \ref{tab:diff3} of Appendix
\ref{difftables}, and the singularities and residues of the Pad\'e
approximants for $D=2,3$ in Tables \ref{tab:pade2} and \ref{tab:pade3}
of Appendix \ref{padetables}.
Pad\'e approximants are of little use in
determining the singularities of a series unless we have reason to
expect the series to be well represented by a rational function. As we
have no reason to expect this to be the case for the virial series of
hard spheres we do not make reference to these tables below, but
include them for the interested reader to compare with the work of
other authors~\cite{ree1964a,ree1967a,jansevanrensburg1993a,sanchez1994a}.
We have not included tables for $D=4,\cdots,8$ for reasons of length,
and feel that the brief summary we make of the information in each of
the tables is sufficient. The dedicated reader may obtain the tables
for singularities and exponents/residues in all dimensions, as well as
tables of coefficients predicted by the approximants, either by
downloading them~\cite{clisbyweb} or upon request from the authors.

Some care needs to be taken in interpreting the tables of
 singularities and exponents from the differential approximants, and
 we
 now discuss some of the relevant issues. Defective
approximants, which have a singularity with an anomalously small
exponent are marked with the symbol $\dag$, and are somewhat
arbitrarily chosen as it is not possible to rigorously define what is
an anomalously small exponent. We use the {\em ad hoc} procedure of
 Hunter and Baker~\cite{hunter1973a}, outlined by
 Guttmann~\cite{guttmann1989a}, that approximants be deemed defective
 if they have a pole with residue less than 0.003 that lies
 approximately inside 
 the radius of convergence of the series. 
 Defective
 approximants should be avoided as they are really lower order
 approximants in disguise, with an extra factor from a denominator
 pole and zero in the numerator that almost cancel. One other
 observation that should be made is that we expect in general if a
 series in the variable $z$ has a singularity $z_c$ with exponent
 $\alpha$, then the approximants may be expected to determine $z_c$
 with more accuracy than $\alpha$. Thus we expect more scatter between
 approximants in 
 exponent values than in the position of singularities. In
 particular, even though we make an observation below about the
 position of the leading singularity in \tab{tab:diff3} for $D=3$,
 nothing can be said about the exponent. Further evidence for the
 unreliability of the exponents comes in the form of the large
 imaginary part of many of the exponents, which is
 very unlikely to be a true feature of the series.
See Guttmann~\cite{guttmann1989a}
for a much deeper discussion on both of the issues raised
 above. 

We have combined the coefficients predicted by the various
approximants~\cite{clisbyweb} to produce Table \ref{tab:predicted}
of predicted coefficients for dimensions $D=2,\cdots,8$. We discuss
these predicted coefficients as  well as 
information obtained about singularities from the differential
approximants for dimensions $D=2,\cdots,8$ below. 

\subsubsection{Analysis}
\label{diffanalysis}

For $D=2$, the leading singularities of the differential approximants lie on
the positive real axis with the exceptions of $[3,3;1]$ and $[3,2,2;0]$.
 In most cases the singularity is located close to
$B_2\rho=1.98$, with an exponent of $-1.74(3)$. We list the
singularities of the approximants in \tab{tab:diff2} of Appendix
\ref{difftables}. 
To make this
estimate we have looked at all approximants which were not defective
that use all 10 terms in 
the series, and calculated an average value. We discarded the values
from the approximant $[2,2,2;0]$ which appeared to be an outlier. No other
singularities follow a consistent pattern, and thus we see no signs of
singularities that could eventually lead to sign changes in the virial
series. As can be seen~\cite{clisbyweb},
predicted coefficients are stable to $B_{18}$, and are listed 
in \tab{tab:predicted} where the scatter in the predicted coefficient
is in the last digit. The Pad\'e approximants
$[4/5]$ and $[5/4]$ may be used as approximate equations of
state at low density, and although they may be obtained directly from
the data we include them for reference purposes along with the
compilation of tables~\cite{clisbyweb}. 

\footnotesize
\be
f_{[4/5]}(\rho) =     \frac{1
  + 0.69939247(B_2\rho)
  - 0.33033017(B_2\rho)^2 
  + 0.11294457(B_2\rho)^3 
  - 0.012320562(B_2\rho)^4  }{
  1
  - 0.30060752(B_2\rho)
  - 0.81172709(B_2\rho)^2
  + 0.62751627(B_2\rho)^3
  - 0.17862580(B_2\rho)^4   
  + 0.021359218(B_2\rho)^5 }
\ee
\be
f_{[5/4]}(\rho) =     \frac{1
   -0.062894522(B_2\rho)
  + 0.13851476(B_2\rho)^2 
  + 0.0067699403(B_2\rho)^3 
  + 0.0039942056(B_2\rho)^4   
  + 0.00047760798(B_2\rho)^5  }{
  1
 -  1.0628945(B_2\rho)
 +  0.41940485(B_2\rho)^2
 -  0.11367848 (B_2\rho)^3
 +  0.021846467(B_2\rho)^4 }
\ee
\normalsize

The variation in the polynomial coefficients in these
approximants 
when one takes in to account the uncertainty in the virial
coefficients is of the same order as the coefficients
themselves. Hence one should not ascribe too much importance to the exact
value of the Pad\'e coefficients as they will change as future improvements
are made in the accuracy of the virial coefficients.

For $D=3$, the position of the leading singularity varies more than
for $D=2$: in most cases the leading singularities for the high order
differential approximants are in the complex
plane with negative real part. However, the position and nature of
this singularity varies substantially between approximants, and it is
not possible to make a definitive statement as to whether this is a
true singularity of the series, or an artifact of the approximant
method. For approximants which only allow for 2 singular points,  the
leading singularity is on the real axis, 
and may be positive or negative. There does however appear in all
approximants to be a
stable singularity on the positive real axis, which is generally not
the leading singularity. It is located at $B_2\rho=3.75(3)$, with an
exponent of $-2.10(8)$, with the estimate made from all high order,
non-defective differential approximants in
\tab{tab:diff3} of Appendix \ref{difftables}. In accordance with the discussion
of Subsection \ref{ratio} this is the singularity that dominates the
series initially. If the complex conjugate pair of singularities are
indeed the leading singularities of the series then this will
eventually lead to a change in sign of the virial series in 3
dimensions. This can be confirmed or rejected by calculating more
coefficients, and possibly by increasing the accuracy of the
coefficients to $B_{10}$ which may reduce the scatter in the position
of the singularities. As can be seen~\cite{clisbyweb},
predicted coefficients are stable to $B_{16}$,
and estimates of their values are given
in \tab{tab:predicted}. We explicitly give here the Pad\'e approximants
$[4/5]$ and $[5/4]$ which may be used as approximate equations of
state at low density.

\footnotesize
\be
f_{[4/5]}(\rho) =     \frac{1   +
0.50998996(B_2\rho)         + 0.20874890
(B_2\rho)^2        + 0.036450422  (B_2\rho)^3       +
0.0035176485 (B_2\rho)^4 }{
        1     -0.49001003(B_2\rho)        + 0.073758936
	(B_2\rho)^2 
	-0.018001749(B_2\rho)^3
	+0.0057761992(B_2\rho)^4
	-0.00054759070(B_2\rho)^5}
\ee
\be
f_{[5/4]}(\rho) =     \frac{        1   +
  0.80408617 (B_2\rho)      +
  0.46662045(B_2\rho)^2    +
  0.14197965(B_2\rho)^3    +
  0.029738518(B_2\rho)^4    +
  0.0028192477(B_2\rho)^5  }{
  1
 -0.19591382(B_2\rho)
 +0.037534279(B_2\rho)^2
 -0.060057986(B_2\rho)^3
 +0.012302957(B_2\rho)^4 }
\ee
\normalsize

As is the case for $D=2$, if one takes in to account the uncertainty
of the virial coefficients then the variance of the coefficients in
the Pad\'e approximants is of the same order as the coefficients themselves.

In $D=4$ there is no clear behavior for the leading singularity that
can be gleaned  from the approximants; there are cases where the
singularity is on the positive real axis, on the negative real axis,
or in the complex plane. Predicted coefficients are correspondingly
variable, and this may be attributed to the low magnitude and
correspondingly poor accuracy of
$B_9$ and especially $B_{10}$. Almost all approximants predict that
$B_{11}$ will be
positive (see \tab{tab:predicted}),
 but after that the predictions diverge: some allow for
negative coefficients as soon as $B_{12}$, and many allow for negative
$B_{13}$ or $B_{14}$, but this is in no way an improvement on our
qualitatively based prediction in Subsection \ref{ratio} that we will
soon see negative coefficients for $D=4$. Higher order virial
coefficients and 
better accuracy for the known virial coefficients will help to resolve
this question.

For $D=5$ the approximants give no clue as to the position or nature
of the leading 
singularity. Predicted coefficients are stable only to $B_{12}$, and
are given in \tab{tab:predicted}. Some approximants predict negative
coefficients for $B_{13}$, and there are no predictions of asymptotic
behavior that are consistent between the different approximants.

We will discuss the cases of $D=6,7,8$ together, as they are all very
similar. The leading singularity is generally on the negative real
axis, which suggests that the alternation in sign of the virial
coefficients will continue for higher orders. There is no apparent
pattern in the position of the next to leading singularity. Predicted
coefficients are stable to $B_{16}$ for $D=6$, and to $B_{18}$ for
$D=7,8$.  This behavior is consistent with either the leading
singularity being on the negative real axis as the differential
approximant analysis suggests, or that there are complex conjugate
singularities adjacent to the negative real axis but close enough so
that there position cannot be accurately determined with ten virial
coefficients. 

\begin{table}[H]
\centering
\caption[Predicted coefficients for approximants with 10 exact
  coefficients in dimensions $D=2,\cdots,8$.]{Predicted coefficients
  for approximants with 10 exact coefficients in dimensions
  $D=2,\cdots,8$, with uncertainty in the last digit.}
\label{tab:predicted}
\vspace{2ex}
\scriptsize
\begin{tabular}{ccccccccc}
\tableline
\tableline
\\[-1.5ex]
\multicolumn{9}{c}{Predicted coefficients}\\
 $D$ &\multicolumn{1}{c}{$B_{11}/B_2^{10}$} &\multicolumn{1}{c}{$B_{12}/B_2^{11}$} &\multicolumn{1}{c}{$B_{13}/B_2^{12}$} &\multicolumn{1}{c}{$B_{14}/B_2^{13}$} &\multicolumn{1}{c}{$B_{15}/B_2^{14}$} &\multicolumn{1}{c}{$B_{16}/B_2^{15}$} &\multicolumn{1}{c}{$B_{17}/B_2^{16}$} &\multicolumn{1}{c}{$B_{18}/B_2^{17}$}\\[0.7ex]
\tableline
\\[-1.5ex]
2 &$ 1.089 \times 10^{-2}$ &$ 5.90 \times 10^{-3}$ &$ 3.18 \times 10^{-3}$ &$ 1.70 \times 10^{-3}$ &$ 9.10 \times 10^{-4}$ &$ 4.84 \times 10^{-4}$ &$ 2.56 \times 10^{-4}$ &$ 1.36 \times 10^{-4}$ \\
3 &$ 1.22 \times 10^{-4}$ &$ 3.64 \times 10^{-5}$ &$ 1.08 \times 10^{-5}$ &$ 3.2 \times 10^{-6}$ &$ 9.2 \times 10^{-7}$ &$ 2.6 \times 10^{-7}$ &&\\
4 &$ 1.2 \times 10^{-6}$&&&&&&&\\
5 &$ 3 \times 10^{-5}$&$-4\times 10^{-5}$&&&&&&\\
6 &$ 4.32 \times 10^{-4}$ &$-3.68 \times 10^{-4}$ &$ 3.2 \times 10^{-4}$ &$-2.9 \times 10^{-4}$ &$ 2.7 \times 10^{-4}$ &$-2.5 \times 10^{-4}$ &&\\
7 &$ 1.43 \times 10^{-3}$ &$-1.40 \times 10^{-3}$ &$ 1.42 \times 10^{-3}$ &$-1.45 \times 10^{-3}$ &$ 1.5 \times 10^{-3}$ &$-1.6 \times 10^{-3}$ &$ 1.8 \times 10^{-3}$ &$-2.1 \times 10^{-3}$ \\
8 &$ 2.56 \times 10^{-3}$ &$-2.72 \times 10^{-3}$ &$ 2.97 \times 10^{-3}$ &$-3.4 \times 10^{-3}$ &$ 4.0 \times 10^{-3}$ &$-4.4 \times 10^{-3}$ &$ 5.5 \times 10^{-3}$ &$-6 \times 10^{-3}$ \\[0.8ex]
\tableline
\end{tabular}
\normalsize
\end{table}

\section{Conclusion}
\label{conclusion}

The key result is that negative virial coefficients have been found
for $D\ge 5$, and a
strong signal was found via the method of ratio analysis in Subsection
\ref{ratio} that negative coefficients will soon occur for
$D=4$. On the basis of evidence from the methods
of ratio analysis in 
Subsection \ref{ratio} and differential approximants in Subsection
\ref{diff} there is a distinct possibility that at higher order some
virial coefficients for $D=3$ will be negative. This would contradict
the frequent
assumption that all virial coefficients are positive for hard spheres
in three
dimensions. More work needs
to be done to confirm or reject this hypothesis; in particular
the calculation of $B_{11}$ and possibly $B_{12}$ may provide the
necessary evidence. For $D=2$ no deviation is seen from the behavior
of a series that is dominated by a singularity on the positive axis at
a density close to the space filling density. However, as discussed in
Subsection \ref{zero}, unless there is good reason to believe that the
series has entered the asymptotic regime then it is impossible to draw
strong conclusions about the position and nature of the leading singularity.

\vspace{0.2cm}

\noindent {\bf Acknowledgments:} {This work was supported in part by 
the National Science Foundation under DMR-0302758. N.~Clisby
gratefully acknowledges support from the Australian Research Council.}

\appendix

\section{Number of Configurations}
\label{spantable}

Numerical results for the virial coefficients are to be found in
\tab{tab:numericalvirial}. 
We list the number of batches of $10^7$ configurations in
\tab{tab:batches}. The total number of
configurations can be obtained by multiplying the number of batches by the
number of spanning trees used at a particular order.

\begin{table}[!htb]
\centering
\caption[Batches used for virial coefficient calculations]{\centering
  Number of batches of $10^7$ configurations used in virial
  coefficient calculations, as a function of order and dimension}
\label{tab:batches}
\vspace{2ex}
\begin{tabular}{rrrrrrrr}
\tableline
\tableline
\\[-1.5ex]
\multicolumn{1}{c}{$D$} 
&\multicolumn{1}{c}{$B_4/B_2^3$} &\multicolumn{1}{c}{$B_5/B_2^4$}
&\multicolumn{1}{c}{$B_6/B_2^5$} &\multicolumn{1}{c}{$B_7/B_2^6$}
&\multicolumn{1}{c}{$B_8/B_2^7$} &\multicolumn{1}{c}{$B_9/B_2^8$}
&\multicolumn{1}{c}{$B_{10}/B_2^9$} \\[0.7ex]
\tableline
\\[-1.5ex]
2& $ 1000 $& $ 9625 $& $ 9000 $& $ 9000 $& $ 15384 $& $ 19553 $& $ 6149 $\\
3& $ 1000 $& $ 52573 $& $ 53463 $& $ 63751 $& $ 64675 $& $ 87609 $& $ 151349 $\\
4& $ 1000 $& $ 8454 $& $ 9400 $& $ 10299 $& $ 21400 $& $ 31903 $& $ 38699 $\\
5& $ 23199 $& $ 8436 $& $ 8618 $& $ 8597 $& $ 15607 $& $ 21042 $& $ 15398 $\\
6& $ 1000 $& $ 8423 $& $ 8600 $& $ 8542 $& $ 5899 $& $ 6300 $& $ 1229 $\\
7& $ 23010 $& $ 8213 $& $ 8600 $& $ 8500 $& $ 5898 $& $ 6300 $& $ 1300 $\\
8& $ 1000 $& $ 8209 $& $ 8500 $& $ 8493 $& $ 5763 $& $ 6265 $& $ 1300 $\\[0.8ex]\tableline
\end{tabular}
\normalsize
\end{table}

\section{Star Content Results}
\label{histograms}

Here we plot histograms of the star content for labeled graphs
contributing to $B_6$, $B_7$, $B_8$, $B_9$, and $B_{10}$.

\begin{figure}[H]
\centering
\includegraphics[scale=0.45,origin=c,angle=-90]{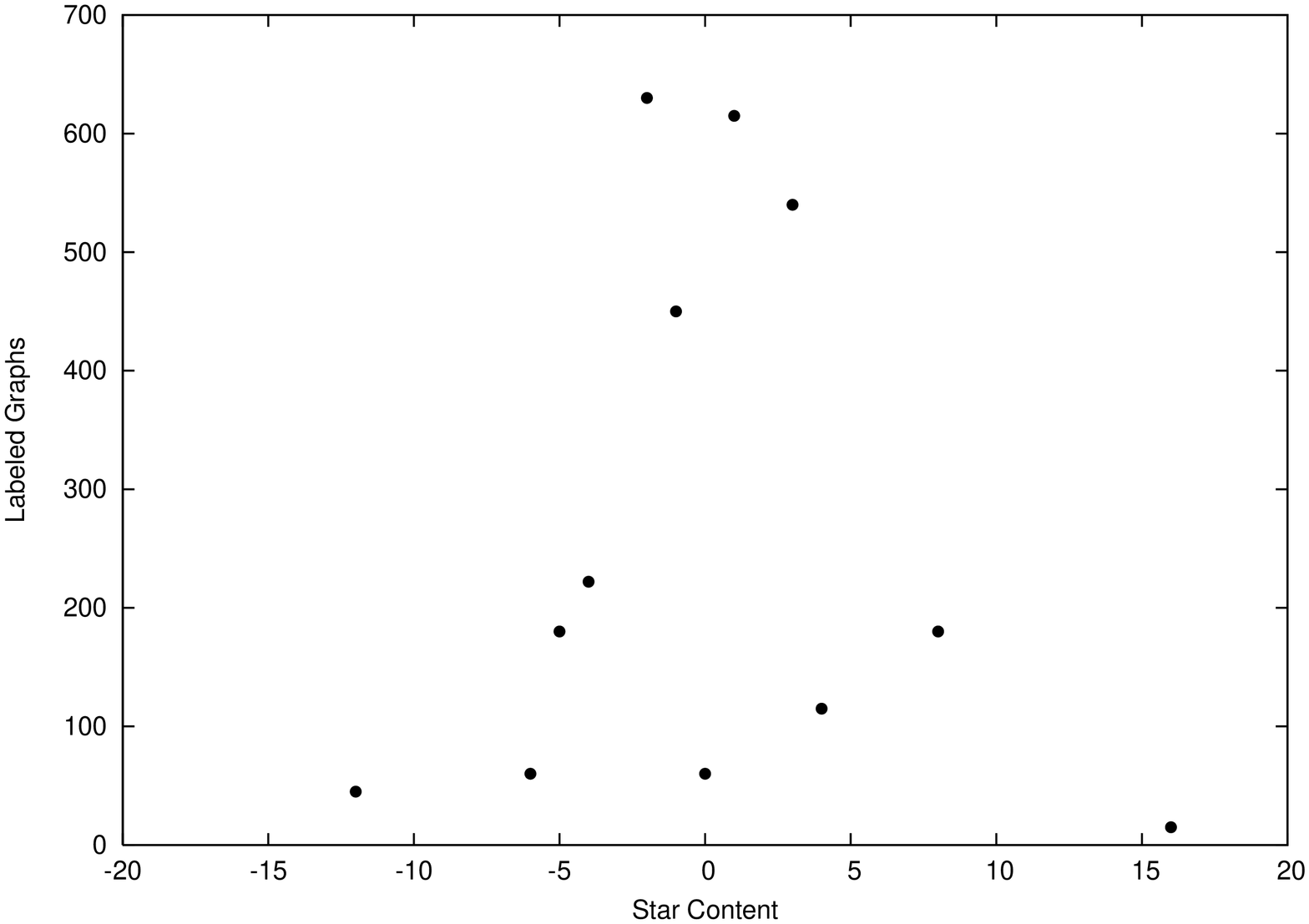}
\vspace{-2cm}
\caption[Star content histogram for $B_6$]{\centering Histogram plot of number of labeled graphs
  versus star content for $B_6$.}
\label{graph:star6}
\end{figure}

\begin{figure}[H]
\centering
\includegraphics[scale=0.45,origin=c,angle=-90]{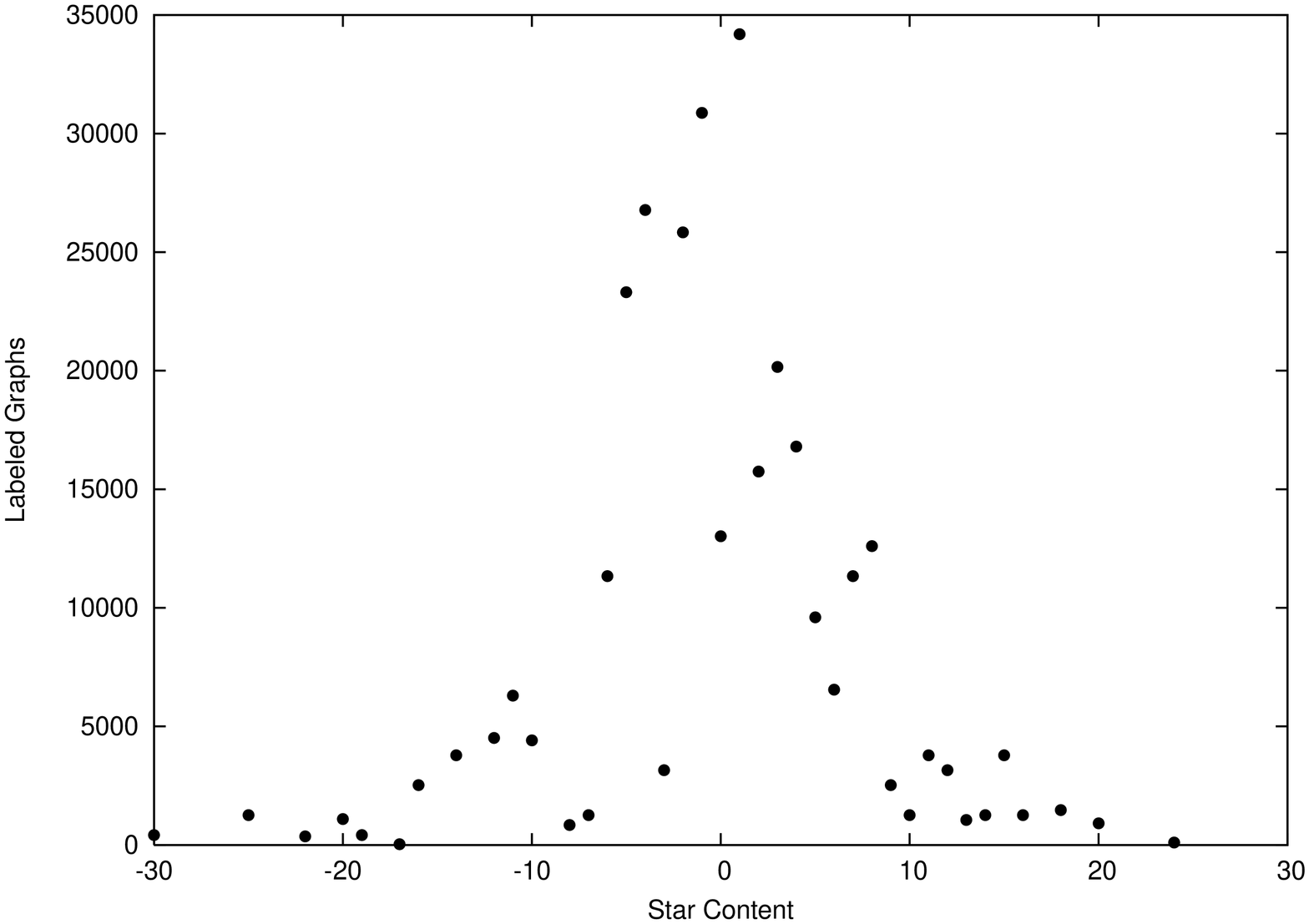}
\vspace{-2cm}
\caption[Star content histogram for $B_7$]{\centering Histogram plot of number of labeled graphs
  versus star content for $B_7$.}
\label{graph:star7}
\end{figure}

\begin{figure}[H]
\centering
\includegraphics[scale=0.45,origin=c,angle=-90]{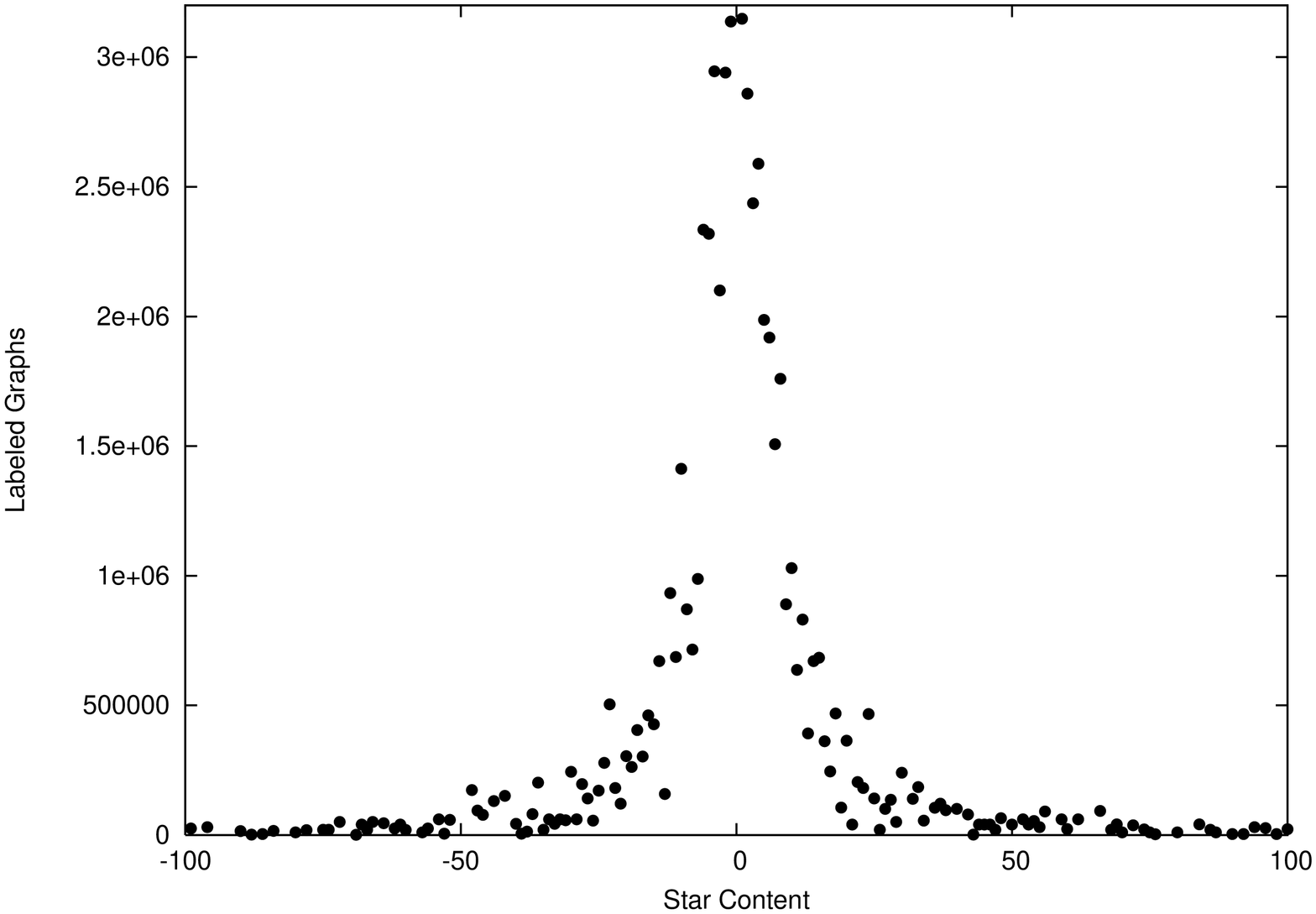}
\vspace{-2cm}
\caption[Star content histogram for $B_8$]{\centering Histogram plot of number of labeled graphs
  versus star content for $B_8$.}
\label{graph:star8}
\end{figure}

\begin{figure}[H]
\centering
\includegraphics[scale=0.45,origin=c,angle=-90]{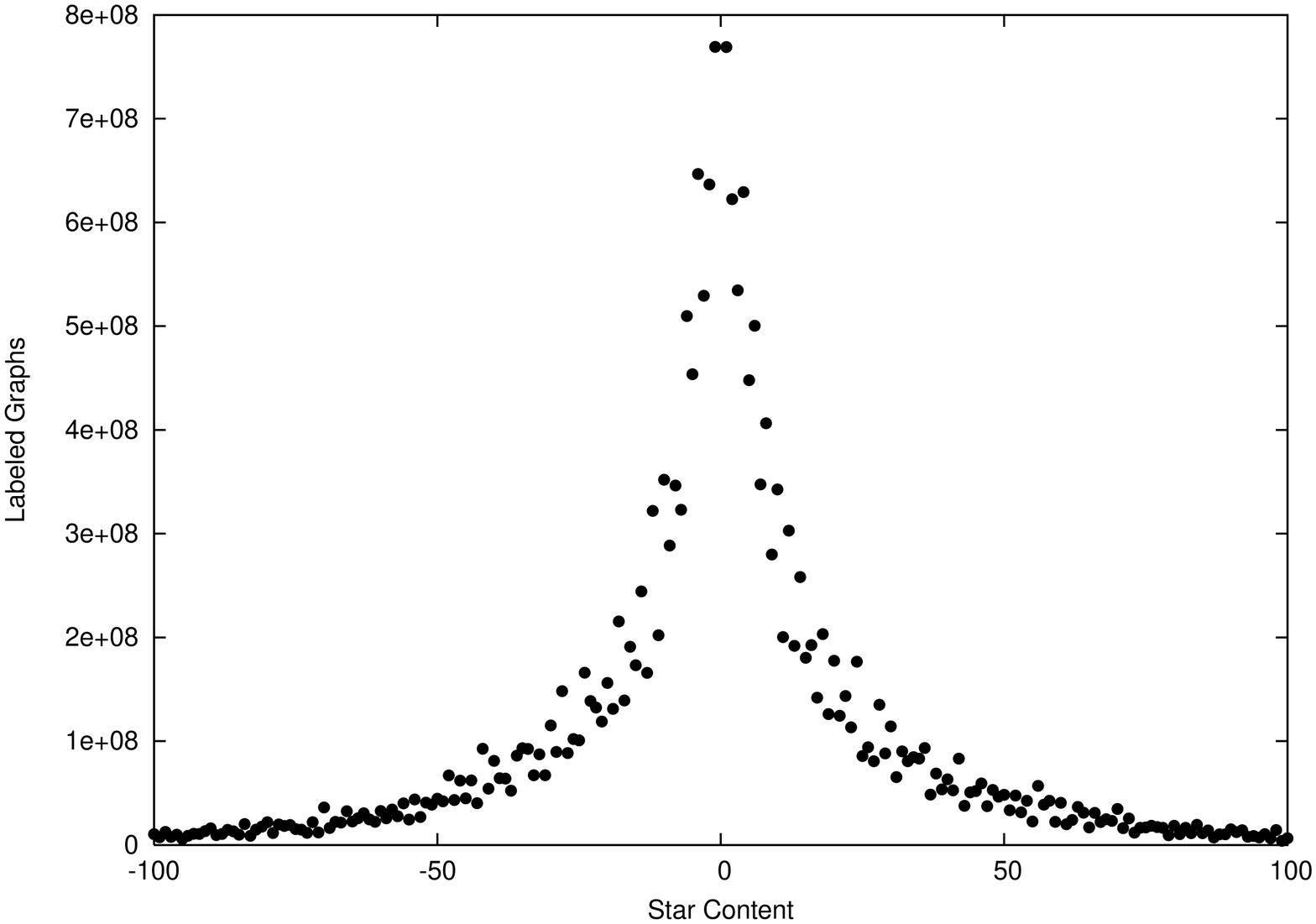}
\vspace{-2cm}
\caption[Star content histogram for $B_9$]{\centering Histogram plot of number of labeled graphs
  versus star content for $B_9$.}
\label{graph:star9}
\end{figure}

\begin{figure}[H]
\centering
\includegraphics[scale=0.45,origin=c,angle=-90]{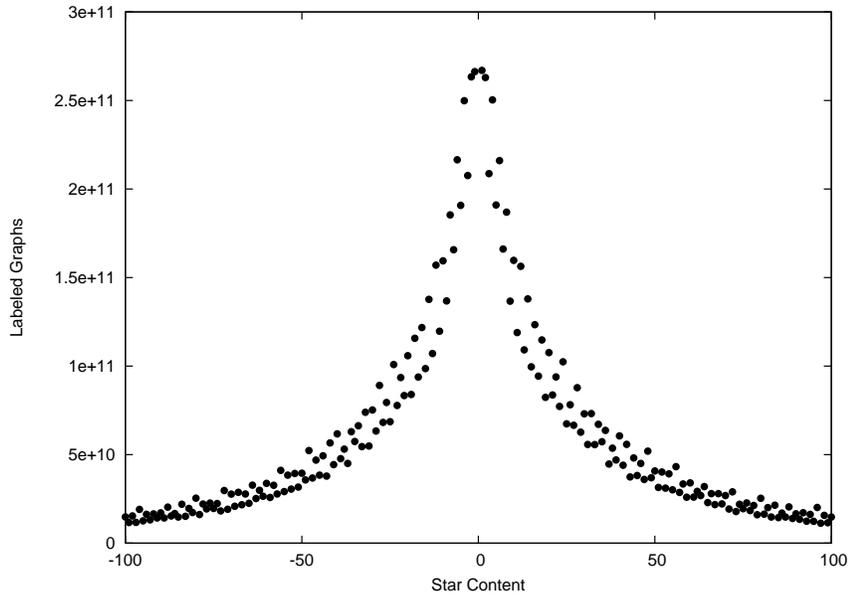}
\vspace{-2cm}
\caption[Star content histogram for $B_{10}$]{\centering Histogram plot of number of labeled graphs
  versus star content for $B_{10}$.}
\label{graph:star10}
\end{figure}

\section{Spanning trees used in the Monte-Carlo Algorithm}
\label{spanfig}

\begin{figure}[H]
\centering
\includegraphics[width=9cm]{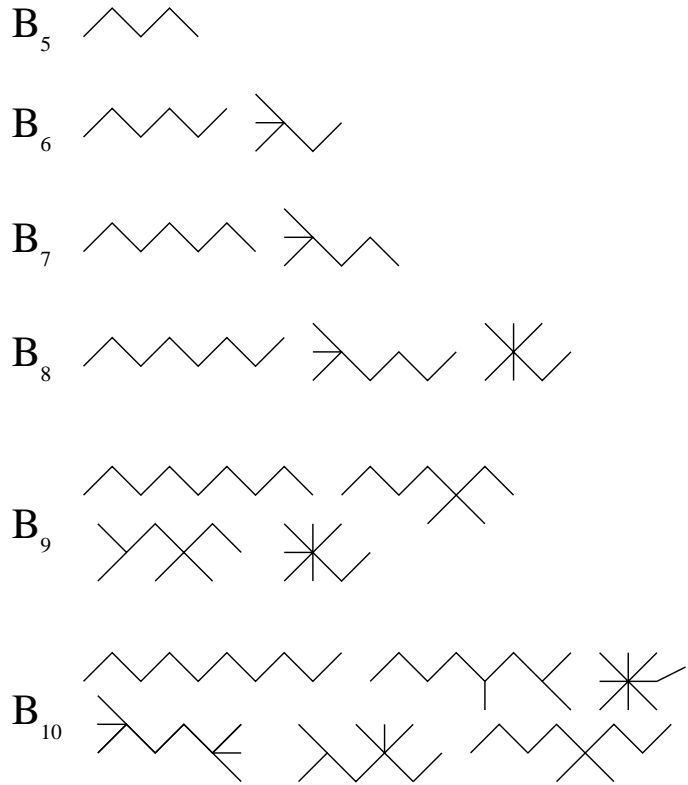}
\caption[Minimal sets of spanning trees for $B_k$,
  $k=5,\cdots,10$]{\centering Minimal sets of spanning trees for $B_k$, $k=5,\cdots,10$.}
\label{fig:minspan}
\end{figure}

\section{Ratio Plots}

\begin{figure}[H]
\centering
\includegraphics[scale=0.45,origin=c,angle=-90]{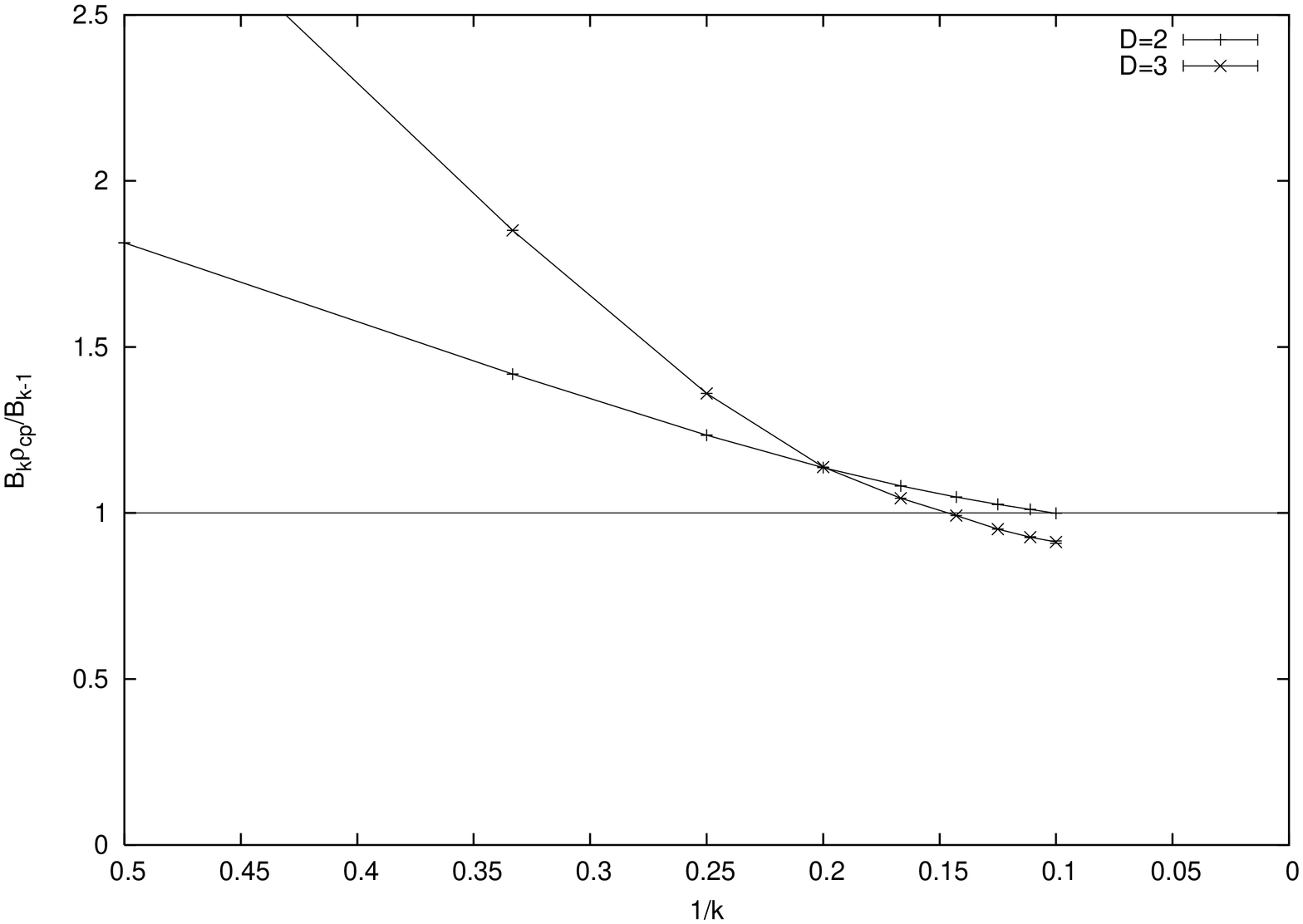}
\vspace{-2cm}
\caption[Ratio plot for virial coefficients in dimensions
 $D=2,3$]{\centering Ratio plot for
virial coefficients in dimensions $D=2,3$.}
\label{graph:ratio1}
\end{figure}

\begin{figure}[H]
\centering
\includegraphics[scale=0.45,origin=c,angle=-90]{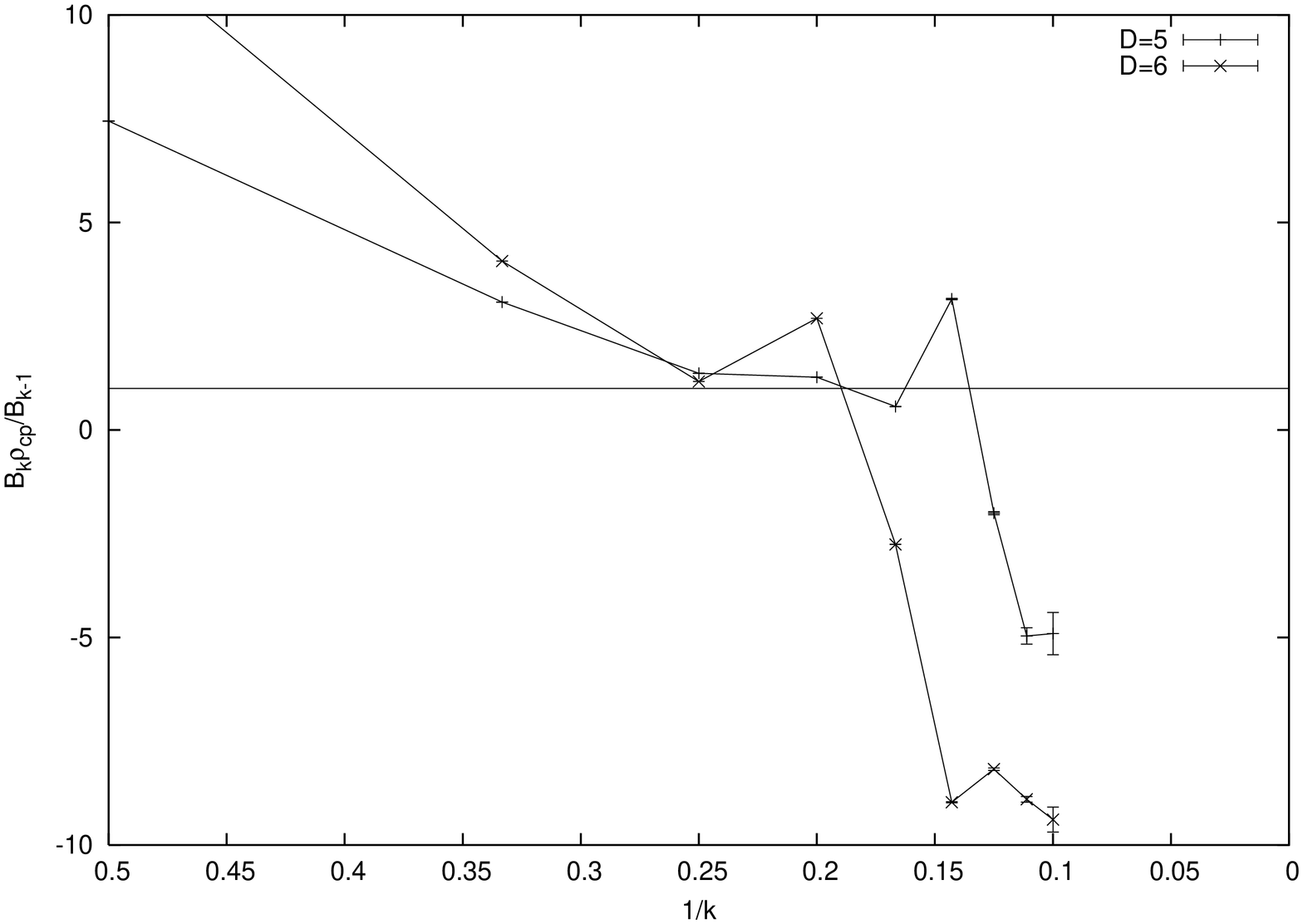}
\vspace{-2cm}
\caption[Ratio plot for virial coefficients in dimensions
 $D=5,6$]{\centering Ratio plot for
virial coefficients in dimensions $D=5,6$.}
\label{graph:ratio2}
\end{figure}

\begin{figure}[H]
\centering
\includegraphics[scale=0.45,origin=c,angle=-90]{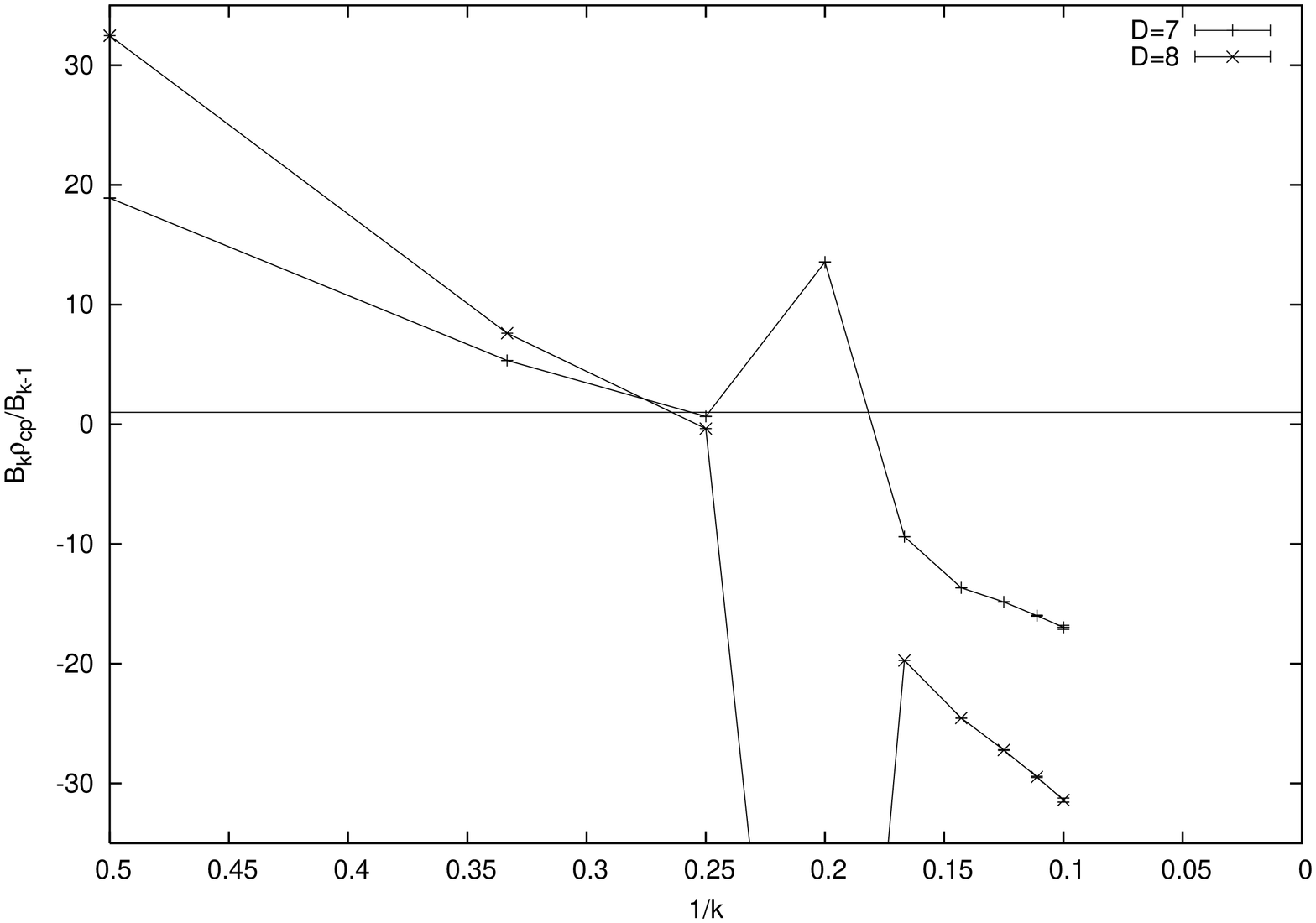}
\vspace{-2cm}
\caption[Ratio plot for virial coefficients in dimensions
 $D=7,8$]{\centering Ratio plot for
virial coefficients in dimensions $D=7,8$.}
\label{graph:ratio3}
\end{figure}

\begin{figure}[H]
\centering
\includegraphics[scale=0.45,origin=c,angle=-90]{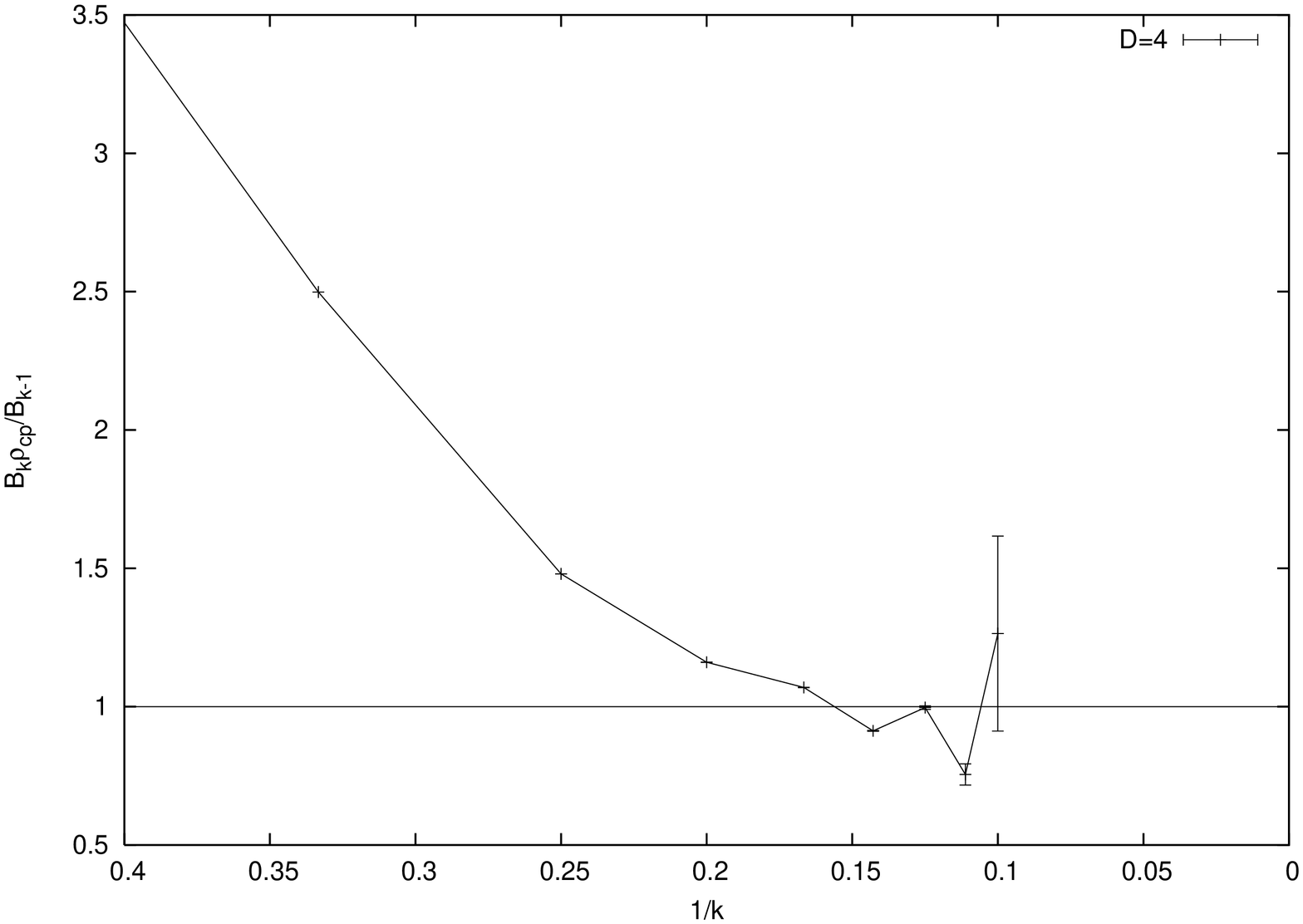}
\vspace{-2cm}
\caption[Ratio plot for virial coefficients in dimension
 $D=4$]{\centering Ratio plot for
virial coefficients in dimension $D=4$.}
\label{graph:ratio4}
\end{figure}

\begin{figure}[H]
\centering
\includegraphics[scale=0.45,origin=c,angle=-90]{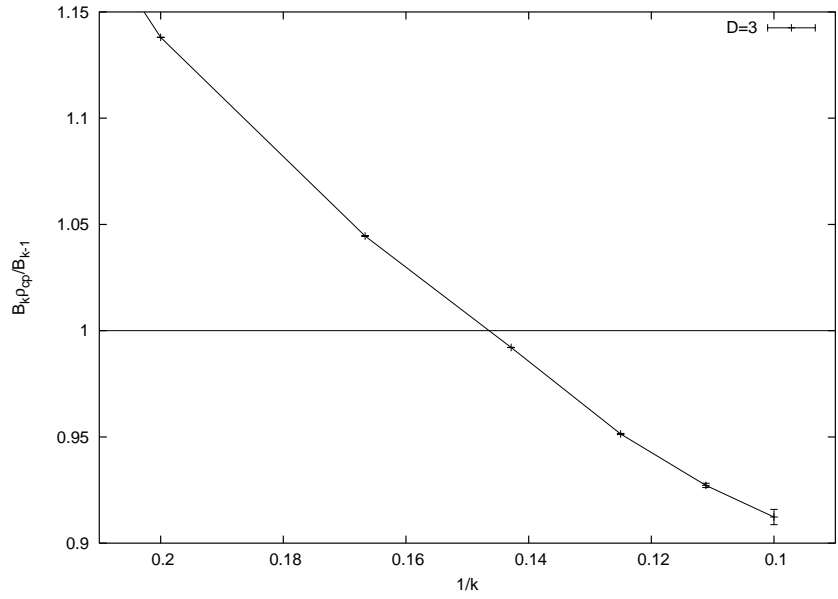}
\vspace{-2cm}
\caption[Ratio plot for virial coefficients in dimensions
 $D=3$]{\centering Ratio plot for
virial coefficients in dimension $D=3$ over a small domain to show
 non-monotonic behavior of the second derivative.}
\label{graph:ratio5}
\end{figure}

\section{Differential Approximants}
\label{difftables}

\begin{table}[H]
\centering
\caption{Singularities for all differential
  approximants in $D=2$ in terms of $B_2\rho$, with the corresponding
  exponents immediately below. Defective approximants are marked with
  $\dag$, and the singularities are arranged so that the singularity
  nearest $B_2\rho = 1.98$ is in the left most column. In most cases, this singularity has the smallest modulus, but when this is not the case the singularity with smallest modulus is
  marked with $*$.}
\label{tab:diff2}
\vspace{2ex}
\small
\begin{tabular}{crrrr}
\tableline
\tableline
\\[-1.5ex]
Approximant &\multicolumn{4}{c}{Singularity / Exponent}\\[0.7ex]
\tableline
\\[-1.5ex]
$[4,3;0]$ &$      1.98$ &$      1.58$ $\pm$ $      2.94$ $\; i$ &  & \\
  &$     -1.74$ &$     0.528$ $\mp$ $     0.449$ $\; i$ &  & \\[0.5ex]
$[3,4;0]$ &$      1.98$ &$      1.58$ $\pm$ $      2.92$ $\; i$ &$     -30.6$ & \\
  &$     -1.74$ &$     0.508$ $\mp$ $     0.453$ $\; i$ &$     0.728$ & \\[0.5ex]
$[4,4;0]^\dag$ &$      1.99$ &$      1.46$ $\pm$ $      2.80$ $\; i$ &$     -1.11^*$ & \\
   &$     -1.79$ &$     0.406$ $\mp$ $     0.386$ $\; i$ &$  3.18 \times 10^{-5}$& \\[0.5ex]
$[5,4;0]^\dag$ &$      1.99$ &$      1.46$ $\pm$ $      2.81$ $\; i$ &$    -0.978^*$ & \\
   &$     -1.79$ &$     0.418$ $\mp$ $     0.390$ $\; i$ &$  1.38 \times 10^{-5}$& \\[0.5ex]
$[4,5;0]^\dag$ &$      1.99$ &$      1.46$ $\pm$ $      2.81$ $\; i$ &$     -267.$&$    -0.977^*$ \\
   &$     -1.79$ &$     0.418$ $\mp$ $     0.390$ $\; i$ &$     0.952$&$  1.37 \times 10^{-5}$\\[0.5ex]
$[3,2;1]$ &$      1.96$ &$     -5.22$ &  & \\
  &$     -1.66$ &$      12.3$ &  & \\[0.5ex]
$[2,3;1]$ &$      1.96$ &$     -3.66$ $\pm$ $      7.70$ $\; i$ &  & \\
  &$     -1.67$ &$      1.92$ $\mp$ $      7.72$ $\; i$ &  & \\[0.5ex]
$[3,3;1]$ &$      1.80$ $\pm$ $     0.198$ $\; i$ &$      3.77$ &  & \\
  &$     -1.07$ $\mp$ $    0.0804$ $\; i$ &$     -2.70$ &  & \\[0.5ex]
$[4,3;1]$  &$      1.98$ &$     -2.58$ &$    -0.601^*$& \\
   &$     -1.74$ &$     -6.19$ &$      27.3$& \\[0.5ex]
$[3,4;1]$ &$      1.98$ &$    -0.589$ $\pm$ $      4.37$ $\; i$ &$    -0.828^*$ & \\
   &$     -1.76$ &$    0.0841$ $\mp$ $      2.37$ $\; i$ &$      8.79$& \\[0.5ex]
$[2,2;2]$ &$      1.96$ &$     -5.78$ &  & \\
  &$     -1.66$ &$      14.0$ &  & \\[0.5ex]
$[3,2;2]$  &$      1.95$ &$     0.619^*$&  & \\
   &$     -1.59$ &$    -0.935$&  & \\[0.5ex]
$[2,3;2]$ &$      2.06$ &$      2.09$ $\pm$ $     0.978$ $\; i$ &  & \\
  &$     -2.25$ &$    -0.830$ $\pm$ $      1.11$ $\; i$ &  & \\[0.5ex]
$[3,3;2]$ &$      1.97$ &$     0.945$ $\pm$ $      1.80$ $\; i$ &  & \\
  &$     -1.70$ &$    -0.344$ $\pm$ $      2.38$ $\; i$ &  & \\[0.5ex]
$[2,2;3]$  &$      1.98$ &$     0.345^*$&  & \\
   &$     -1.79$ &$     -23.7$&  & \\[0.5ex]
$[3,2;3]$  &$      1.98$ &$     0.361^*$&  & \\
   &$     -1.78$ &$     -21.2$&  & \\[0.5ex]
$[2,3;3]$ &$      1.98$ &$      42.6$ &$     0.445^*$ & \\
   &$     -1.78$ &$     -3.23$ &$     -16.6$& \\[0.5ex]
$[2,2;4]$  &$      1.98$ &$   -0.0937^*$  & \\
   &$     -1.75$ &$      95.9$&  & \\[0.5ex]
$[2,2,2;0]$  &$      1.79$ &$      1.21$&  & \\
   &$      1.24$ &$     -10.7$&  & \\[0.5ex]
$[3,2,2;0]$ &$      2.16$ $\pm$ $     0.319$ $\; i$ &  &  & \\
  &$     -2.74$ $\mp$ $     0.955$ $\; i$ &  &  & \\[0.5ex]
$[2,3,3;0]$  &$      1.94$ &$     0.939$ $\pm$ $      1.61$ $\; i^*$&  & \\
   &$     -1.41$ &$    -0.985$ $\pm$ $      1.12$ $\; i$&  & \\[0.5ex]
$[2,2,2;1]$ &$      1.98$ &$      9.17$ &  & \\
  &$     -1.72$ &$     -17.5$ &  & \\[0.8ex]
\tableline
\end{tabular}
\normalsize
\end{table}

\begin{table}[H]
\centering
\caption{Singularities for all differential approximants
  in $D=3$ in terms of $B_2\rho$, with the corresponding 
  exponents immediately below. Defective approximants are marked with
  $\dag$, and the singularities are listed from left to right in order 
  of their modulus. The most stable singularity is on the positive real axis in the vicinity of $B_2\rho = 3.75$, and this appears in the second column in all cases.}
\label{tab:diff3}
\vspace{2ex}
\small
\begin{tabular}{crrr}
\tableline
\tableline
\\[-1.5ex]
Approximant &\multicolumn{3}{c}{Singularity / Exponent}\\[0.7ex]
\tableline
\\[-1.5ex]
$[4,3;0]$ &$     -1.03$ $\pm$ $      2.64$ $\; i$ &$      3.71$ & \\
  &$     0.640$ $\mp$ $    0.0898$ $\; i$ &$     -2.04$ & \\[0.5ex]
$[3,4;0]$ &$     -1.05$ $\pm$ $      2.73$ $\; i$ &$      3.83$ &$     -6.75$\\
  &$     0.752$ $\mp$ $     0.134$ $\; i$ &$     -2.33$ &$     0.824$\\[0.5ex]
$[4,4;0]$ &$     -1.04$ $\pm$ $      2.65$ $\; i$ &$      3.73$ &$     -65.6$\\
  &$     0.652$ $\mp$ $    0.0885$ $\; i$ &$     -2.09$ &$      14.1$\\[0.5ex]
$[5,4;0]$ &$     -1.04$ $\pm$ $      2.65$ $\; i$ &$      3.73$ &$     -232.$\\
  &$     0.651$ $\mp$ $    0.0881$ $\; i$ &$     -2.09$ &$      183.$\\[0.5ex]
$[4,5;0]$ &$     -1.04$ $\pm$ $      2.65$ $\; i$ &$      3.73$ &$     -44.4$ $\pm$ $      58.5$ $\; i$\\
  &$     0.651$ $\mp$ $    0.0881$ $\; i$ &$     -2.09$ &$     0.394$ $\mp$ $      9.61$ $\; i$\\[0.5ex]
$[3,2;1]$ &$     -3.49$ &$      4.04$ & \\
  &$      6.71$ &$     -2.95$ & \\[0.5ex]
$[2,3;1]$ &$     -1.68$ $\pm$ $      1.51$ $\; i$ &$      3.79$ & \\
  &$      1.72$ $\mp$ $     0.928$ $\; i$ &$     -2.25$ & \\[0.5ex]
$[3,3;1]$ &$     -1.93$ $\pm$ $      1.76$ $\; i$ &$      3.81$ & \\
  &$      1.54$ $\mp$ $      1.36$ $\; i$ &$     -2.30$ & \\[0.5ex]
$[4,3;1]$ &$     -1.11$ $\pm$ $      2.69$ $\; i$ &$      3.73$ & \\
  &$     0.627$ $\mp$ $     0.190$ $\; i$ &$     -2.07$ & \\[0.5ex]
$[3,4;1]$ &$    -0.621$ $\pm$ $      1.87$ $\; i$ &$      3.79$ &$     -6.95$\\
  &$     0.913$ $\pm$ $      1.18$ $\; i$ &$     -2.24$ &$    -0.506$\\[0.5ex]
$[2,2;2]$ &$      1.15$ &$      3.64$ & \\
  &$     -3.22$ &$     -1.99$ & \\[0.5ex]
$[3,2;2]$ &$    -0.544$ &$      3.87$ & \\
  &$      13.4$ &$     -2.47$ & \\[0.5ex]
$[2,3;2]$ &$     -1.24$ $\pm$ $      2.13$ $\; i$ &$      3.80$ & \\
  &$      1.13$ $\pm$ $     0.230$ $\; i$ &$     -2.27$ & \\[0.5ex]
$[3,3;2]$ &$    -0.464$ $\pm$ $      2.25$ $\; i$ &$      3.78$ & \\
  &$     0.476$ $\pm$ $      1.21$ $\; i$ &$     -2.22$ & \\[0.5ex]
$[2,2;3]$ &$      2.91$ &$      3.75$ & \\
  &$     -1.17$ &$     -1.96$ & \\[0.5ex]
$[3,2;3]$ &$      2.74$ &$      3.77$ & \\
  &$     -1.44$ &$     -2.01$ & \\[0.5ex]
$[2,3;3]$ &$      2.37$ &$      3.77$ &$     -38.7$\\
  &$     -2.05$ &$     -2.09$ &$     -3.71$\\[0.5ex]
$[2,2;4]$ &$      2.78$ &$      3.77$ & \\
  &$     -1.38$ &$     -2.00$ & \\[0.5ex]
$[2,2,2;0]$ &$      4.35$ &$     -5.19$ & \\
  &$     -3.86$ &$      4.07$ & \\[0.5ex]
$[3,2,2;0]$ &$      1.21$ &$      6.04$ & \\
  &$     -1.35$ &$     -9.45$ & \\[0.5ex]
$[2,3,3;0]$ &$    -0.840$ $\pm$ $      2.45$ $\; i$ &$      3.76$ & \\
  &$     0.469$ $\pm$ $    0.0986$ $\; i$ &$     -2.17$ & \\[0.5ex]
$[2,2,2;1]$ &$     -2.68$ &$      3.70$ & \\
  &$      6.49$ &$     -2.06$ & \\[0.8ex]
\tableline
\end{tabular}
\normalsize
\end{table}

\section{Pad\'e Approximants}
\label{padetables}

\begin{table}[H]
\centering
\caption{Singularities for all Pad\'e approximants in $D=2$ in terms
  of $B_2\rho$, with the corresponding residues immediately below.
 Defective approximants are marked with $\dag$.}
\label{tab:pade2}
\vspace{2ex}
\begin{tabular}{crrrr}
\tableline
\tableline
\\[-1.5ex]
Approximant &\multicolumn{4}{c}{Singularity / Residue}\\[0.7ex]
\tableline
\\[-1.5ex]
$[4/3]$ &$      1.87$ &$      2.40$ &$      19.9$ & \\
  &$     -11.5$ &$      13.7$ &$     -63.2$ & \\[0.5ex]
$[3/4]$ &$      1.91$ &$      2.21$ &$      2.78$ $\pm$ $      2.62$ $\; i$ & \\
  &$     -16.8$ &$      17.4$ &$    -0.568$ $\pm$ $     0.424$ $\; i$ & \\[0.5ex]
$[4/4]$ &$      1.89$ $\pm$ $     0.187$ $\; i$ &$      2.81$ $\pm$ $      1.43$ $\; i$ &  & \\
  &$     -1.33$ $\mp$ $      6.26$ $\; i$ &$     0.839$ $\pm$ $      1.90$ $\; i$ &  & \\[0.5ex]
$[5/4]$ &$      1.90$ &$      2.28$ &$     0.513$ $\pm$ $      3.21$ $\; i$ & \\
  &$     -14.4$ &$      15.8$ &$    0.0116$ $\mp$ $    0.0580$ $\; i$ & \\[0.5ex]
$[4/5]^\dag$ &$    -0.908$ &$      1.94$ &$      2.12$ &$      2.60$ $\pm$ $      2.40$ $\; i$\\
  &$ -4.83 \times 10^{-7}$ &$     -26.8$ &$      27.0$ &$    -0.380$ $\pm$ $     0.551$ $\; i$\\[0.8ex]
\tableline
\end{tabular}
\normalsize
\end{table}

\begin{table}[H]
\centering
\caption{Singularities for all Pad\'e
  approximants in $D=3$ in terms of $B_2\rho$, with the corresponding
  residues immediately below. Defective approximants are marked with
  $\dag$.} 
\label{tab:pade3}
\vspace{2ex}
\begin{tabular}{crrrr}
\tableline
\tableline
\\[-1.5ex]
Approximant &\multicolumn{4}{c}{Singularity / Residue}\\[0.7ex]
\tableline
\\[-1.5ex]
$[4/3]^\dag$ &$     0.516$ &$      3.44$ &$      3.92$ & \\
  &$  2.84 \times 10^{-7}$ &$     -146.$ &$      178.$ & \\[0.5ex]
$[3/4]$ &$      3.32$ $\pm$ $     0.602$ $\; i$ &$     0.249$ $\pm$ $      7.66$ $\; i$ &  & \\
  &$      7.42$ $\mp$ $      36.9$ $\; i$ &$     -1.87$ $\mp$ $      5.40$ $\; i$ &  & \\[0.5ex]
$[4/4]$ &$     -2.65$ &$      3.50$ $\pm$ $     0.461$ $\; i$ &$     -16.3$ & \\
  &$    0.0100$ &$      11.9$ $\mp$ $      65.0$ $\; i$ &$     -37.2$ & \\[0.5ex]
$[5/4]$ &$     -1.13$ $\pm$ $      2.24$ $\; i$ &$      3.57$ $\pm$ $     0.366$ $\; i$ &  & \\
  &$   0.00100$ $\mp$ $   0.00669$ $\; i$ &$      14.0$ $\mp$ $      92.3$ $\; i$ &  & \\[0.5ex]
$[4/5]$ &$      3.43$ &$     -1.80$ $\pm$ $      3.99$ $\; i$ &$      4.39$ &$      6.32$\\
  &$     -109.$ &$     0.135$ $\mp$ $     0.293$ $\; i$ &$      213.$ &$     -110.$\\[0.8ex]
\tableline
\end{tabular}
\normalsize
\end{table}

\end{document}